\documentclass[reprint, nobibnotes, amsmath,amssymb, aps, prd, floatfix]{revtex4-2}
\usepackage[T1]{fontenc} 
\usepackage{graphicx} 
\usepackage{dcolumn} 
\usepackage{bm} 
\usepackage{xcolor} 
\usepackage{ulem} 
\usepackage{textcomp} 
\usepackage{hyperref}

\DeclareMathOperator{\sech}{sech}
\newcounter{AssumptionListCounter}  
\newcommand{\Msat}{m}
\newcommand{\MMW}{M}
\newcommand{\Msun}{M_\odot}
\newcommand{\rscr}{r_\mathrm{scr}}
\newcommand{\rscrMW}{r_\mathrm{scr,MW}}
\newcommand{\rscrsat}{r_\mathrm{scr,sat}}
\newcommand{\rscrmax}{r_\mathrm{scr,max}}
\newcommand{\Qsat}{Q_\mathrm{sat}}
\newcommand{\tE}{\tilde{E}}

\newcommand{\rt}{r_\mathrm{t}}
\newcommand{\xt}{x_\mathrm{t}}

\newcommand{\aS}{a_\mathrm{S}}

\def\xs{{\mathbf r_{\rm s}}}
\def\xcm{{\mathbf r_{\rm cm}}}
\def\modxs{{r_{\rm s}}}
\def\xsdd{{\mathbf {\ddot r}_{\rm s}}}
\def\xsd{{\mathbf {\dot r}_{\rm s}}}
\def\xh{{\mathbf r_{\rm h}}}
\def\modxh{{r_{\rm h}}}
\def\rh{\modxh}
\def\Omegas{{\mathbf \Omega_{\rm s}}}
\def\modOmegas{{\Omega_{\rm s}}}
\def\Omegab{{\mathbf \Omega}}
\def\rt{r_{\rm t}}

\begin{document}

\title{Stellar Streams in Chameleon Gravity}

\author{A. P. Naik}
\email{an485@ast.cam.ac.uk}
\author{N. W. Evans}
\email{nwe@ast.cam.ac.uk}
\affiliation{Institute of Astronomy, University of Cambridge, Madingley Road, Cambridge, CB3 0HA, UK}

\author{E. Puchwein}
\affiliation{Leibniz-Institut f\"ur Astrophysik Potsdam, An der Sternwarte 16, 14482 Potsdam, Germany}

\author{H. Zhao}
\email{hz4@st-andrews.ac.uk}
\affiliation{Kavli Institute for Cosmology Cambridge, Madingley Road, Cambridge, CB3 0HA, UK}
\affiliation{Scottish Universities Physics Alliance, University of St Andrews, North Haugh, St Andrews, Fife KY16 9SS, UK}

\author{A.-C. Davis}
\affiliation{Department of Applied Mathematics and Theoretical Physics, Centre for Mathematical Sciences, Cambridge CB2 0WA, UK}

\date{\today}

\begin{abstract}
Theories of gravity that incorporate new scalar degrees of freedom typically require `screening mechanisms' to ensure consistency with Solar System tests. One widely-studied mechanism---the chameleon mechanism---can lead to violations of the equivalence principle (EP), as screened and unscreened objects fall differently. If the stars are screened but the surrounding dark matter is not, EP-violation can lead to asymmetry between leading and trailing streams from tidally disrupted dwarf galaxies in the Milky Way halo. We provide analytic estimates of the magnitude of this effect for realistic Galactic mass distributions, demonstrating that it is an even more sensitive probe than suggested previously. Using a restricted N-body code, we simulate 4 satellites with a range of masses and orbits, together with a variety of strengths of the fifth force and screening levels of the Milky Way and satellite. The ratio of the cumulative number function of stars in the leading and trailing stream as a function of longitude from the satellite is computable from simulations, measurable from the stellar data and can provide a direct test of chameleon gravity. We forecast constraints for streams at large Galactocentric distances, which probe deeper into chameleon parameter space, using the specific example case of Hu-Sawicki $f(R)$ gravity. Streams in the outer reaches of the Milky Way halo (with apocentres between 100 and 200 kpc) provide easily attainable constraints at the level of $|f_{R0}| = 10^{-7}$. Still more stringent constraints at the level of $10^{-7.5}$ or even $10^{-8}$ are plausible provided the environmental screening of the satellite is accounted for, and screening of the Milky Way's outer halo by the Local Group is not yet triggered in this range. These would be among the tightest  astrophysical constraints to date. We note three further signatures of chameleon gravity: (i) the trailing stellar stream may become detached from the dark matter progenitor if all the stars are lost, (ii) in the extreme fifth force regime, striations in the stellar trailing tail may develop from material liberated at successive pericentric passages, (iii) if the satellite is fully screened, its orbital frequency is lower than that of the associated dark matter, which is preferentially liberated into the leading tidal tail.
\end{abstract}

\maketitle


\section{Introduction}
\label{S:Introduction}

Stellar streams and substructures are the wreckage of dwarf galaxies and globular clusters that have fallen into and are being torn apart by the Milky Way's tidal field. In the past, such substructures have usually been identified as over-densities of resolved stars, as in the  `Field of Streams' image from the Sloan Digital Sky Survey \citep{Belokurov2006}. There, using multi-band photometry, the stellar halo of the Milky Way was revealed as being composed of criss-crossing stellar streams, the detritus of satellite galaxies. However, streams and substructure remain kinematically cold and so identifiable in phase space long after they have ceased to be recognisable in star counts against the stellar background of the Galaxy \citep{Johnston1998}. The debris persists for a large fraction of a Hubble time, sometimes longer, so substructures in phase space remain to the present day. Searches in phase space for streams are much more powerful than searches in configuration space.

The Gaia satellite is a scanning satellite of the European Space Agency that is monitoring all objects brighter than magnitude $G \approx 20$ around 70 times over a period of 5 years (though the mission lifetime has recently been extended) \citep{Gaia2016, Gaia2018}. Its telescopes are providing magnitudes, parallaxes, proper motions and broad band colours for over a billion stars in the Galaxy ($\approx 1$ per cent of the Milky Way stellar population) within the Gaia-sphere -- or within roughly 20 kpc of the Sun for main sequence stars, 100 kpc for giants. We now possess detailed phase space information, often with spectroscopic and chemical data from cross-matches with other surveys. This has led to the discovery of abundant streams and substructures \citep{Myeong2018,Malhan2018,Meingast2019,Koposov2019}. Streams discovered by Gaia are already being followed up spectroscopically to give six-dimensional (6D) phase space data \citep{Li2019}. Bright tracers such as blue horizontal branch stars or RR Lyraes can be seen out to distances of 250 kpc, assuming Gaia's limiting magnitude of G ~ 20.5. Stars near the tip of the red giant branch can be seen even further out to at least 600 kpc. In future, this should enable Gaia to provide astrometry for very distant streams, perhaps beyond the edge of the Milky Way's dark halo.

If a stream were a simple orbit, then the positions and velocities of stars would permit the acceleration and force field to be derived directly from the 6D data. Streams are more complex than orbits \citep{Sanders2013,Bowden2015}, but the principle remains the same -- their evolution constrains the matter distribution and theory of gravity. Although this idea has been in the literature for some years, exploitation has been sparse primarily because of the limited number of streams with 6D data before Gaia. This field is therefore ripe for further exploitation in the Gaia Era.

Because of their different ages and different positions in phase space, different streams may tell us different things about the theory of gravity. For example, \citet{Thomas2018} show that streams from globular clusters are lopsided in Modified Newtonian Dynamics or MOND because the `external field effect' violates the strong equivalence principle. Meanwhile, \citet{Kesden2006a,Kesden2006b} demonstrated that if a so-called `fifth force' couples to dark matter but not to baryons, this violation of the equivalence principle (EP) leads to large, observable asymmetries in stellar streams from dark matter dominated dwarf galaxies. Specifically, the preponderance of stars are disrupted via the outer Lagrange point rather than the inner one, and the trailing stream is consequently significantly more populated than the leading one. Building on that work, \citet{Keselman2009} explored the regime of fifth forces much stronger than those investigated by Kesden and Kamionkowski and found a number of interesting results, including plausible formation scenarios for the Sagittarius stream, the Draco satellite, and progenitor-less `orphan' streams around the Milky Way.

In the intervening years since the work of Kesden and Kamionkowski, screened modified gravity theories have become the subject of increasing attention~\citep{Amendola2010, Clifton2012, Joyce2015, Koyama2016}. In these theories, a scalar field coupled to gravity is introduced, giving rise to gravitational-strength `fifth forces'. For the field to retain cosmological relevance while also avoiding violations of stringent Solar System tests of gravity, `screening mechanisms' are introduced \citep{Jain2010, Khoury2010}. There are several varieties of screening mechanism, but in the one studied here---the chameleon mechanism---the mass of the scalar field is environment-dependent, such that the fifth force is suppressed within deep potential wells \citep{Khoury2004}. In other words, in dense environments like our Solar System, the chameleon becomes invisible to fifth force searches, hence its name.

A widely-studied class of modified gravity theories is $f(R)$ gravity \citep{Buchdahl1970}. Here, the Ricci scalar $R$ in the Einstein-Hilbert action is generalised to $R+f(R)$. The Hu-Sawicki form of $f(R)$ \citep{Hu2007} exhibits the chameleon mechanism and has been shown to be formally equivalent to a subclass of scalar-tensor theories of gravity \citep{Brax2008}. The key parameter is the present-day cosmic background value of the scalar field, $f_{R0}$. In the present work, we do not assume Hu-Sawicki $f(R)$ gravity, but sometimes use the parameter $f_{R0}$ as a concrete example to illustrate the possible constraints achievable from stellar streams, noting that constraints are also obtainable in the wider chameleon space.

A complete compendium of current constraints on $f(R)$ gravity and chameleon gravity more generally can be found in the review article by \citet{Burrage2018}. It is worth noting that some of the strongest constraints to date have come from weak-field astrophysical probes. Moreover, \citet{Baker2019} identify a `desert' in modified gravity parameter space accessible only to galaxy-scale probes, and have launched the `Novel Probes' project, aimed at connecting theorists with observers in order to devise tests to probe this region. Accordingly, several recent works \citep{Naik2018, Naik2019, Desmond2018a, Desmond2018b, Desmond2019, Vikram2018} have studied imprints of screened modified gravity on galaxy scales.

In chameleon theories, main sequence stars will have sufficiently deep potential wells to self-screen against the fifth force. A diffuse dark matter or gaseous component of sufficiently low mass, however, will be unscreened. As a result, the EP is effectively violated, leading to a number of distinct signatures, as listed by \citet{Hui2009}. Indeed, several of the galaxy-scale studies mentioned in the previous paragraph searched for signatures in this list, as well as other signatures of EP-violation.

The present work explores the idea that effective EP-violation of chameleon gravity should give rise to the stellar stream asymmetries predicted by \citet{Kesden2006a,Kesden2006b}. We will show that tidal streams in the Milky Way, observable with Gaia, can provide constraints that are comparable to, or stronger than, other astrophysical probes. Section \ref{S:Theory} gives a brief introduction to the fifth force in chameleon theories before providing a new calculation of the magnitude of the effect, extending the original work of \citet{Kesden2006a}.
Next, Section \ref{S:Models} describes the Milky Way and satellite models that we use in our simulation code, the methodology and validation of which are in turn described respectively in Sections \ref{S:Methods} and \ref{S:CodeValidation}. Section \ref{S:Results} describes results for a range of tidal streams, inspired by examples discovered recently in large photometric surveys or the Gaia datasets. Finally, Section \ref{S:Conclusions} gives some concluding remarks.

\section{Theory}
\label{S:Theory}

\begin{figure*}[htp]
    \centering
    \includegraphics{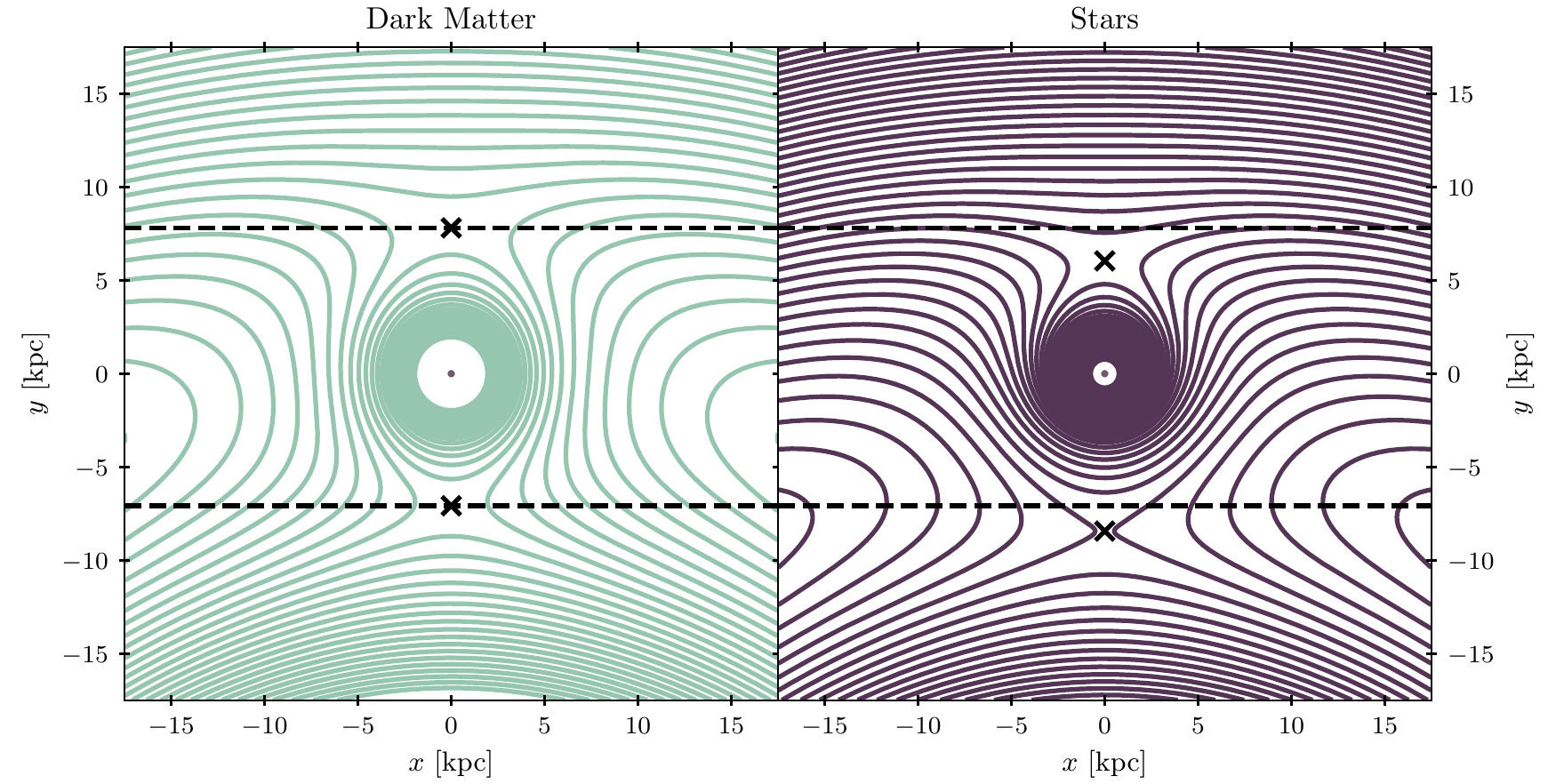}
    \caption{\textit{Left:} Contour map of the effective potential for the dark matter $\Phi_\mathrm{eff, DM}$ as given by Eq.~(\ref{E:PhiEffDM}). \textit{Right:} Contour map of the effective potential for the stars $\Phi_\mathrm{eff, *}$ as given by Eq.~(\ref{E:PhiEffStars}). In both panels, the satellite is at the origin and the Galactic centre is at $(0, -50, 0)$ kpc. The inner and outer Lagrange points are marked by crosses. To guide the eye, black dashed lines marking the positions of the DM Lagrange points span the figure. The asymmetry of the Lagrange points for the stellar effective potential illustrates the cause of the stream asymmetries under chameleon gravity. (Parameters: $\MMW=10^{12}M_\odot, \Msat=10^{10}M_\odot$, and $\beta=0.5$.)}
    \label{F:PhiEffective}
\end{figure*}

\subsection{Chameleon Fifth Forces}
\label{S:Theory:FifthForce}

In scalar-tensor gravity theories, new scalar degrees of freedom in the gravitational sector couple to matter, giving rise to gravitational strength `fifth forces'. Chameleon theories are a class of scalar-tensor theories in which these fifth forces are suppressed in regions of high-density or deep gravitational potential \citep{Khoury2004}. In this section, we cover only the most salient aspects of chameleon theories, and refer the reader elsewhere \cite{Burrage2018} for a more complete description of the formalism and summary of existing constraints.

Consider a spherical overdensity embedded within a region of average cosmic density. If the gravitational well of the object is sufficiently deep, a central region of radius $\rscr$ will be `screened', such that no fifth forces act within the region; $\rscr$ is the `screening radius' of the object.  Outside the screening radius, an unscreened test particle will experience an acceleration due to the fifth force given by Eq.~(3.6) of Ref~\cite{Burrage2018}
\begin{equation}
\label{E:FifthForce}
    a_5(r) = 2 \beta^2 \frac{G\left(M(r)-M(\rscr)\right)}{r^2},
\end{equation}
where $M(r)$ is the mass enclosed within radius $r$, and $\beta$ is the coupling strength. In other words, the fifth force is sourced only by the mass lying between the screening radius and the test particle. We have also assumed here that the Compton wavelength of the theory is much larger than the characteristic length scales of the system.

Eq.~(\ref{E:FifthForce}) gives the fifth force experienced by an unscreened test particle outside the screening radius of an overdense object, but the situation is complicated further in the case where instead of a test particle, we have another extended object -- for example, a star or dwarf galaxy situated outside the screening radius of its host galaxy. In this case, the acceleration of object $i$ (mass $M_i$, radius $r_i$, screening radius $r_{\mathrm{scr},i}$) due to the fifth force is given by
\begin{equation}
\label{E:FifthForceExtendedParticle}
    a_5(r) = 2 \beta^2 Q_i \frac{G\left(M(r)-M(\rscr)\right)}{r^2},
\end{equation}
where $Q_i$ is the `scalar charge' of object $i$, given in turn by
\begin{equation}
\label{E:ScalarCharge}
    Q_i = \left(1-\frac{M_i(r_{\mathrm{scr}, i})}{M_i}\right).
\end{equation}
Thus, the test object experiences the full fifth force only if it is fully unscreened (i.e., $\rscr=0$) and experiences no fifth force if it is fully screened ($\rscr=r_i$). In the intermediate case where the object is partially screened, it experiences a reduced fifth force. In this work, we assume stars to be fully self-screened (i.e. $Q=0$), and dark matter to be fully unscreened (i.e. $Q=1$). For the satellite galaxies (i.e. the stream progenitors), we explore a number of regimes, spanning fully screened, partially screened, and fully unscreened.

A commonly used parametrisation of chameleon theories is in terms of the coupling strength $\beta$ and the `self-screening parameter' $\chi_c$. The latter parameter determines which astrophysical objects are fully or partially screened, and can be used to calculate their screening radii. Note that in the case of Hu-Sawicki $f(R)$ gravity, $\beta$ is fixed to $\sqrt{1/6}$, while $\chi_c=-f_{R0}$.

In order to derive constraints in the $\beta/\chi_c$ plane or $f_{R0}$ space from stellar streams around the Milky Way, we would need to adopt some prescription to convert $\chi_c$ to Milky Way and satellite screening radii. Analytical formulae exist in the case of an isolated spherical body \citep{Davis2012, Sakstein2013}, but such a treatment would neglect the environmental contribution of the Local Group to the Milky Way's screening, the environmental contribution of the Milky Way to the satellite's screening, and the impact of the non-sphericity of the Milky Way. The calculation therefore requires numerical methods in more realistic scenarios \citep{Naik2018}. In this work, we instead use $\beta$, $\rscrMW$, and $\rscrsat$ as input parameters for reasons of computational ease. However, in Section \ref{S:Results:Future}, we investigate the connection between $f_{R0}$ and $\rscrMW$ in order to forecast constraints from future data.

\subsection{Stream Asymmetries}
\label{S:Theory:Asymmetries}

\subsubsection{A Physical Picture}

We begin with a physical picture of the cause of stream asymmetries. Consider a satellite represented by a point mass $\Msat$. For the moment, let us neglect any fifth forces and assume that the Milky Way and can also be represented as a point mass $\MMW$, so both satellite and the Milky Way are moving on circular orbits with frequency $\Omega$ around their common center of mass.

We use a coordinate system whose origin is at the centre of the satellite. Then, a star at position $\xs$ moves in an `effective' gravitational potential given by~\citep[e.g.,][]{Goldstein1950,Binney2008}
\begin{equation}
\label{E:PhiEff}
    \Phi_\mathrm{eff}(\xs) = - \frac{G\Msat}{\modxs} - \frac{G\MMW}{|\xh-\xs|} - \frac{1}{2}\Omega^2 |\xs-\xcm|^2.
\end{equation}
where $\xh$ is the position of the point mass representing the Milky Way and $\xcm$ is the position of the centre of mass. We use the convention $\modxs = |\xs|$ to denote the modulus of any vector. The first two terms are the gravitational potentials of the satellite and Milky Way respectively, while the final term provides the centrifugal force due to the frame of reference, which is rotating about the centre of mass with frequency $\Omega$
\begin{equation}
\label{E:Lagrange}
    \Omega = \sqrt{\frac{G(\MMW+\Msat)}{\modxh^3}}.
\end{equation}
In practice, the mass of a typical satellite $\Msat$ is at least two orders of magnitude less than the mass of the Milky Way, and so its contribution to the frequency can be neglected.

The stationary points of the effective potential $\Phi_\mathrm{eff}$ are the Lagrange points or equilibria at which the net force on a star at rest vanishes. In the circular restricted three-body problem, there are five Lagrange points. Matter is pulled out of the satellite at the `L1' and `L2' saddle points, henceforth the `inner' and `outer' Lagrange points. These are situated either side of the satellite, co-linear with the satellite and Milky Way. Leading (trailing) streams originate at the inner (outer) Lagrange points, which lie at (see Section 8.3.1 of \citet{Binney2008})
\begin{equation}
\label{E:tidal_rad}
    \rt \approx \left(\frac{\Msat}{3\MMW} \right)^{1/3}\modxh,
\end{equation}
with respect to the satellite centre.

Now consider how the system behaves if a fifth force acts on the dark matter. Neglecting any screening, and assuming the satellite is dark matter dominated, the orbit will circle more quickly with frequency given by
\begin{equation}
    \Omega' = \sqrt{\frac{G'(\MMW+\Msat)}{\modxh^3}} \approx \sqrt{\frac{G'\MMW}{\modxh^3}}
\end{equation}
where $G' \equiv (1+2\beta^2)G$.
The effective potential experienced by a dark matter particle in this system is
\begin{equation}
\label{E:PhiEffDM}
    \Phi_\mathrm{eff, DM}(\xs) = - \frac{G'\Msat}{\modxs} - \frac{G'\MMW}{|\xh-\xs|} - \frac{1}{2}\Omega'^2 |\xs-\xcm|^2.
\end{equation}
This is tantamount to a linear rescaling of Eq.~(\ref{E:PhiEff}), and the locations of the critical points are therefore unchanged relative to the standard gravity case. However, the effective potential is different for a star which does not feel the fifth force, namely
\begin{equation}
\label{E:PhiEffStars}
    \Phi_\mathrm{eff, *}(\xs) = - \frac{G\Msat}{\modxs} - \frac{G\MMW}{|\xh-\xs|} - \frac{1}{2}\Omega'^2 |\xs-\xcm|^2.
\end{equation}
This is not a linear multiple of Eq.~(\ref{E:PhiEff}), and the locations of the Lagrange points are consequently altered. The two panels of Figure~\ref{F:PhiEffective} shows contour maps of the effective potentials for dark matter and stars, for $\MMW=10^{12}M_\odot, \Msat=10^{10}M_\odot, \modxh=50$ kpc, and $\beta=0.5$. Also indicated on the diagram are the locations of the inner and outer Lagrange points of the potentials.

In the dark matter case, the points are approximately equidistant from the satellite centre. However, a significant asymmetry is visible in the stellar effective potential, with the outer Lagrange point being much closer to the satellite and at a lower effective potential. Thus, stars are much more likely to be stripped from the satellite at the outer Lagrange point, and the trailing stream will consequently be more populated than the leading one.

Physically, we can understand this effect in terms of force balance. The stars are being dragged along by the satellite, which is orbiting at an enhanced rotation speed due to the fifth force. This enhanced speed means that the outward centrifugal force on the stars is greater than the inward gravitational attraction by the Milky Way. The consequence of this net outward force is that stars can be stripped from the satellite more easily if they are at larger Galactocentric radii than the satellite, and less easily if they are at smaller radii. This is reflected in the positions of the Lagrange points.

Stars unbound from the satellite will be on a slower orbit around the Milky Way than their progenitor. If $\beta$ is sufficiently large, then stars that are initially in the leading stream can fall behind and end in the trailing stream.

\subsubsection{Circular Restricted Three-Body Problem}

We now solve for the stream asymmetries in the circular restricted three-body problem, following and correcting Ref.~\cite{Kesden2006a}. This is a useful preliminary before passing to the general case. In Newtonian gravity, the forces balance at the inner and outer Lagrange points, and so
\begin{align}
\label{E:ForceL1}
-\frac{G\MMW}{(\rh -\rt)^2} + \frac{G\Msat}{\rt^2}
+ \frac{G(\MMW+\Msat)}{\rh^3}\left( \frac{\MMW \rh}{\MMW+\Msat} -\rt\right)=0,\\
\label{E:ForceL2}
-\frac{G\MMW}{(\rh +\rt)^2} - \frac{G\Msat}{\rt^2}
+ \frac{G(\MMW+\Msat)}{\rh^3}\left( \frac{\MMW \rh}{\MMW+\Msat} +\rt\right)=0.
\end{align}
We recall that the inertial frame is rotating about the centre of mass, and so the centrifugal terms in Eqs.~(\ref{E:ForceL1}) and (\ref{E:ForceL2}) depend on the distance of the Lagrange point to the centre of mass, not the Galactic centre (cf. Eqs.~(14) and (15) of Ref.~\cite{Kesden2006a}).

We now define $u = \rt/\modxh$ and $u' = \rt'/\modxh$ for the inner and outer Lagrange points respectively, and obtain
\begin{align}
    u^3 & =  \frac{\Msat}{\MMW}\frac{ (1-u^3)(1-u)^2}{ 3 -3u +u^2},\\
   u'^3 & =  \frac{\Msat}{\MMW}\frac{ (1-u'^3)(1+u')^2}{ 3 +3u' + u'^2 }.
\end{align}
Solving, we find that
\begin{align}
    u & \approx \left( \frac{\Msat}{3\MMW} \right)^{1/3}\left(1 -\frac{u}{3}\right),\\
    u' & \approx \left( \frac{\Msat}{3\MMW} \right)^{1/3} \left( 1 + \frac{u'}{3} \right),
\end{align}
so the natural asymmetry is
\begin{equation}
\label{E:threebodyasym}
\Delta r_{\rm nat} = (u'-u)\rh \approx \frac{2}{3}\left(\frac{\Msat}{3\MMW}\right)^{2/3}\rh.
\end{equation}

Now introducing a fifth force, the force balance equations for stars not directly coupling to the fifth force become
\begin{align}
\label{E:ForceL1F}
-\frac{G\MMW}{(\rh -\rt)^2} + \frac{G\Msat}{\rt^2}
+ \Omega^2(1+2\beta^2)\left( \frac{\MMW \rh}{\MMW+\Msat} -\rt\right)=0,\\
\label{E:ForceL2F}
-\frac{G\MMW}{(\rh +\rt)^2} - \frac{G\Msat}{\rt^2}
+ \Omega^2(1+2\beta^2)\left( \frac{\MMW \rh}{\MMW+\Msat} +\rt\right)=0.
\end{align}
Proceeding as before
\begin{align}
    u & \approx \left( \frac{\Msat}{3\MMW} \right)^{1/3} \frac{1}{(1+2\beta^2)^{1/3}} \left(1 - \frac{u}{3} + \frac{2\beta^2}{3}\frac{\MMW}{\Msat}u^2 \right),\\
   u' & \approx \left( \frac{\Msat}{3\MMW} \right)^{1/3} \frac{1}{(1+2\beta^2)^{1/3}} \left(1 + \frac{u'}{3} - \frac{2\beta^2}{3}\frac{\MMW}{\Msat}u'^2 \right).
\end{align}
The last term on the right-hand side produces an asymmetry with opposite sign to the natural asymmetry. Note that as $u \propto (m/M)^{1/3}$, the $M/m$ factor makes this term actually the largest. The condition for the asymmetry due to the fifth force to overwhelm the Newtonian one is then just
\begin{equation}
\label{E:threebodycrit}
    2\beta^2 \gtrsim 3^{1/3} \left( \frac{\Msat}{\MMW} \right)^{2/3},
\end{equation}
where only leading terms are kept.
This result can be compared with Eq.~(29) of Ref.~\citep{Kesden2006a}. Although the scaling is the same, the numerical factor is different (remember on comparing results that $2\beta^2$ in our paper corresponds to $\beta^2 f_Rf_{\rm sat}$ in theirs). In fact, the changes are very much to the advantage of the fifth force, as smaller values of $\beta$ now give detectable asymmetries.

The two most massive of the MW dwarf spheroidals are Sagittarius with dark matter mass $2.8 \times 10^8 \Msun$ and Fornax at $1.3 \times 10^8 \Msun$ \citep{Amorisco2011}. These will allow values of $\beta^2 \gtrsim 2 \times 10^{-3}$ to be probed. For the smallest dwarf spheroidals such as Segue 1 with a mass of  $6 \times 10^5 \Msun$, then values of $\beta^2 \gtrsim 2 \times 10^{-4}$ are in principle accessible. It should be noted that the Segue 1 is an ambiguous object, and it is not entirely clear if it is a dark matter dominated dwarf or a globular cluster~\citep{Niederste2009}.

\subsubsection{General Case}

The circular restricted three-body problem is somewhat unrealistic, as the Galaxy's matter distribution is extended. In particular, there is a significant difference in the enclosed host mass within the inner and outer Lagrange points and this plays a role in the strength of the asymmetry. We now proceed to give a mathematical analysis of the general case.

The satellite is now moving on a orbit with instantaneous angular frequency $\Omegab$. We work in a (non-inertial) reference frame rotating at $\Omegab$ with origin at the centre of the satellite. A star at location $\xs$ now feels the following forces: (i) a gravitational attraction by the satellite, (ii) a gravitational attraction by the host galaxy; (iii) an inertial force because the satellite is falling into the host and so the reference frame is not inertial and (iv) the Euler, Coriolis and centrifugal forces because the reference frame is rotating. Note that (iii) was not necessary in our earlier treatment of the circular restricted three-body problem because there we chose an inertial frame tied to the centre of mass.

The equation of motion for a star or dark matter particle is
\begin{align}
{\xsdd} & =  -G \Msat(\modxs) \frac{\xs}{\modxs^3}
-G\MMW(|\xs-\xh|)\frac{(\xs-\xh)}{| \xs-\xh |^3}\nonumber \\
&- G\MMW(\modxh)\frac{\xh}{\modxh^3} -{\dot \Omegab} \times \xs  -2 \Omegab \times \xsd \\
& - \Omegab \times( \Omegab \times \xs)\nonumber
\end{align}
Save for the assumption that the matter distributions in the satellite $\Msat(\modxs)$ and the host $\MMW(\modxh)$ are spherically symmetric, this expression is general.

We now assume that the star or dark matter particle is following a circular orbit around the satellite with orbital frequency $\Omegas$ and that $\modxs/\modxh \ll 1$. By careful Taylor expansion, we obtain
\begin{align}
{\xsdd} & =  -G \Msat(\modxs) \frac{\xs}{\modxs^3}
+G\MMW(\modxh)\frac{(3-n)(\xs\cdot\xh) \xh}{| \modxh |^5}\nonumber \\
&- G\MMW(\modxh)\frac{\xs}{\modxh^3} -{\dot \Omegab} \times \xs  -2 \Omegab \times (\Omegas \times \xs) \\
& + \Omegab \times( \Omegab \times \xs)\nonumber
\end{align}
where $n(\modxh)$ is the logarithmic gradient of $\MMW(\modxh)$.

To calculate the tidal radius, we now specialise to the case of a particle whose orbit lies in the same plane as the satellite's orbit. The satellite's circular frequency is $\Omega^2 = GM(\modxh)/\modxh^3$. The tidal radius is defined as the distance from the centre of the satellite at which there is no net acceleration, i.e., the forces on the particle towards the host and the satellite balance. This gives the tidal radius as
\begin{equation}
    \rt =   \frac{1}{(1 - n + 2 \modOmegas/\Omega)^{1/3} } \left( \frac{\Msat(\modxs)}{\MMW(\modxh)} \right)^{1/3} \modxh.
\end{equation}
When satellite and host are point masses, then $n=0$ and $\Omega = \modOmegas$, and we recover the result previously found in Eq.~(\ref{E:tidal_rad}).

We now define $u = \rt/\modxh$ and $u' = \rt'/\modxh$ for the inner and outer Lagrange points respectively, and obtain
\begin{align}
    u^3 & =  \frac{\Msat(\modxs)}{\MMW(\modxh)}\frac{ (1-u)^{2-n} u}{1- (1-u)^{2-n} + \alpha(1-u)^{2-n}u};\\
   u'^3 & =  \frac{\Msat(\modxs)}{\MMW(\modxh)}\frac{ (1+u')^{2-n} u'}{(1+u')^{2-n} -1 + \alpha(1+u')^{2-n}u'},
\end{align}
where $\alpha \equiv 2\modOmegas/\Omega-1$. We now solve for the difference in the positions of the Lagrange points with respect to the satellite centre $u'-u$. This is the natural stream asymmetry
\begin{equation}
\label{E:natasym}
\Delta r_{\rm nat} \approx (u'-u)\rh =\left(\frac{\Msat(\modxs)}{\MMW(\rh)}\right) ^{2/3} \frac{(2-n)(3-n) \rh}{3 (1-n + 2 \modOmegas/\Omega)^{5/3}}
\end{equation}
In the restricted three-body problem, $n=0$ and $\Omega = \modOmegas$, so we recover our previous result in Eq.~(\ref{E:threebodyasym}).

We wish to compare this asymmetry to the asymmetry produced by adding the modified gravity acceleration of the satellite to the equation of motion. Now specialising to the case $\Omega = \modOmegas$ to reduce complexity, we find the asymmetry due to the fifth force is
\begin{equation}
\Delta r_5 \approx -\frac{4}{3(3-n)} \beta^2 \rh.
\end{equation}
So, the requirement that the dark matter asymmetry overwhelms the natural asymmetry is
\begin{equation}
\label{E:gencaseasym}
    2\beta^2 \gtrsim \frac{(2-n)(3-n)}{2(3- n)^{2/3}} \left(\frac{\Msat}{\MMW}\right)^{2/3},
\end{equation}
which again reduces to Eq.~(\ref{E:threebodycrit}) in the restricted three body case, as it should. For galactic dynamics, a reasonable choice is $n=1$, which corresponds to a galaxy with a flat rotation curve, i.e. an isothermal sphere. Assuming the stars in the satellite satisfy $\modOmegas = \Omega$, then tidal streams in galaxies with flat rotation curves are much more sensitive probes of the dark matter asymmetry. As we move from $n=0$ (the point mass case) to $n=1$ (the isothermal sphere), we gain an additional factor of $\approx 2.3$ in sensitivity. The changes are again in our favour. The asymmetries in tidal streams are therefore an even more delicate probe of the fifth force than suggested by the analysis in Ref.~\cite{Kesden2006a}.

\section{Milky Way and Satellite Models}
\label{S:Models}

In our simulations, we follow the evolution of a large number of tracer particles, stripped from a satellite galaxy and forming tidal tails. The test particles are accelerated by the gravity field of both the Milky Way and the satellite, together with any fifth force contributions. We begin by describing our models for the Milky Way and satellite.

\subsection{Milky Way Model}
\label{S:Models:MW}

\begin{table}[htp]
    \caption{Milky Way model parameters from Ref.~\cite{McMillan2017}. The first three columns respectively give the symbol representing a given parameter, the number of the equation in which it appears, and a physical description. The final column lists the parameter values; in most cases these are best-fitting values inferred by Ref.~\cite{McMillan2017}, but some were instead fixed \textit{a priori}, such as the various disc scale heights. Further details can be found in that article.}
    \begin{ruledtabular}
    \begin{tabular}{l l l r l}
    Symbol                    & Eq.             & Parameter                             & \multicolumn{2}{c}{Value}              \\ \hline
    $\rho_{0,h}$              & \ref{E:MWhalo}  & Halo scale density                    & 0.00853702 & $M_\odot/\mathrm{pc}^3$   \\
    $r_{0,h}$                 & \ref{E:MWhalo}  & Halo scale radius                     & 19.5725    & kpc                       \\
    $\rho_{0,b}$              & \ref{E:MWbulge} & Bulge scale density                   & 98.351     & $M_\odot/\mathrm{pc}^3$   \\
    $r_{0,b}$                 & \ref{E:MWbulge} & Bulge scale radius                    & 0.075      & kpc                       \\
    $r_\mathrm{cut}$          & \ref{E:MWbulge} & Bulge cutoff radius                   & 2.1        & kpc                       \\
    $\Sigma_0^\mathrm{thin}$  & \ref{E:MWsdisc} & Thin disc normalisation               & 895.679    & $M_\odot/\mathrm{pc}^2$   \\
    $z_0^\mathrm{thin}$       & \ref{E:MWsdisc} & Thin disc scale height                & 300        & pc                        \\
    $R_0^\mathrm{thin}$       & \ref{E:MWsdisc} & Thin disc scale radius                & 2.49955    & kpc                       \\
    $\Sigma_0^\mathrm{thick}$ & \ref{E:MWsdisc} & Thick disc normalisation              & 183.444    & $M_\odot/\mathrm{pc}^2$   \\
    $z_0^\mathrm{thick}$      & \ref{E:MWsdisc} & Thick disc scale height               & 900        & pc                        \\
    $R_0^\mathrm{thick}$      & \ref{E:MWsdisc} & Thick disc scale radius               & 3.02134    & kpc                       \\
    $\Sigma_0^\mathrm{HI}$    & \ref{E:MWgdisc} & HI disc normalisation                 & 53.1319    & $M_\odot/\mathrm{pc}^2$   \\
    $z_0^\mathrm{HI}$         & \ref{E:MWgdisc} & HI disc scale height                  & 85         & pc                        \\
    $R_0^\mathrm{HI}$         & \ref{E:MWgdisc} & HI disc scale radius                  & 7          & kpc                       \\
    $R_h^\mathrm{HI}$         & \ref{E:MWgdisc} & HI disc hole radius                   & 4          & kpc                       \\
    $\Sigma_0^{\mathrm{H}_2}$ & \ref{E:MWgdisc} & H\textsubscript{2} disc normalisation & 2179.95    & \,$M_\odot/\mathrm{pc}^2$ \\
    $z_0^{\mathrm{H}_2}$      & \ref{E:MWgdisc} & H\textsubscript{2} disc scale height  & 45         & pc                        \\
    $R_0^{\mathrm{H}_2}$      & \ref{E:MWgdisc} & H\textsubscript{2} disc scale radius  & 1.5        & kpc                       \\
    $R_h^{\mathrm{H}_2}$      & \ref{E:MWgdisc} & H\textsubscript{2} disc hole radius   & 12         & kpc                       \\
    \end{tabular}
    \end{ruledtabular}
    \label{T:McMillanPars}
\end{table}

For the Milky Way, we adopt the axisymmetric mass model of Ref.~\cite{McMillan2017}, which is designed to fit a number of recent kinematic constraints on the Milky Way matter distribution. The model comprises six distinct components: a central bulge, a dark matter halo, and four discs (thin and thick stellar discs, and atomic and molecular gas discs). The functional form of the density distribution of each of these components is given as follows. The various undefined symbols are parameters of the model, for all of which we adopt the values of Ref.~\cite{McMillan2017}. We reproduce these values in Table \ref{T:McMillanPars}.

For the DM halo, the model employs a Navarro-Frenk-White (NFW) profile \citep{Navarro1997},
\begin{equation}
\label{E:MWhalo}
    \rho(r) = \frac{\rho_{0,h}}{\left(\frac{r}{r_{0,h}}\right)\left(1+\frac{r}{r_{0,h}}\right)^2}.
\end{equation}
We have experimented with non-spherical oblate and prolate halo profiles, as well as steeper inner slopes (cf. \cite{Schaller2015}), and found no significant impact on our results. However, it would be interesting in future to investigate the impact of a truly triaxial dark matter halo.

Meanwhile, the bulge is represented by an axisymmetrised version of the model of \citet{Bissantz2002},
\begin{equation}
\label{E:MWbulge}
    \rho(R,z) = \frac{\rho_{0,b}}{\left(1+\frac{r'}{r_{0,b}}\right)^{1.8}}e^{-\left(\frac{r'}{r_\mathrm{cut}}\right)^2},
\end{equation}
where $r' \equiv \sqrt{R^2 + 4z^2}$.

The two stellar discs are represented by a simple exponential disc with an exponential vertical profile,
\begin{equation}
\label{E:MWsdisc}
    \rho(R,z) = \frac{\Sigma_0}{2z_0} e^{-R/R_0}e^{-{|z|}/{z_0}},
\end{equation}
while the two gas discs are given by an exponential disc model with a central hole and a `sech-squared' vertical profile,
\begin{equation}
\label{E:MWgdisc}
    \rho(R,z) = \frac{\Sigma_0}{4z_0} e^{-\left(R_h/R + R/R_0\right)}\sech^2\left(\frac{z}{2z_0}\right).
\end{equation}

With this axisymmetric mass model in hand, the gravitational potential is then calculated with a Poisson solver \footnote{\url{https://github.com/aneeshnaik/mw_poisson}} utilising a spherical harmonic technique similar to that described by \citet{Dehnen1998}. The solver calculates the potential and its gradients on a spherical grid; for our simulations we use a grid with 2000 log-spaced radial cells between $10^{-4}$\,kpc $10^{4}$\,kpc, and 2500 (polar) angular cells, and truncate the spherical harmonic expansion at multipole $l=80$. These settings were found to yield converged solutions for the Milky Way potential corresponding to the mass model described above.

The gravitational acceleration on a test particle (neglecting any fifth forces for the moment) due to the Milky Way is then calculated by interpolating the potential gradient at the position of the particle, employing a cubic spline.

\subsection{Satellite Model}
\label{S:Models:Satellite}

We model the satellite with a truncated Hernquist sphere with the density cut off at a radius $\rt$. The reason for this sharp truncation will become clear in the discussion of the fifth force in \S\ref{S:Models:FifthForce}. Defining a reduced radius $x \equiv r/\aS$ (thus $\xt \equiv \rt/\aS$) where $\aS$ is the scale radius of the profile, the density-potential pair is given by
\begin{equation}
\label{E:SatPotential}
    \begin{aligned}
        \Phi(x) &= \begin{cases}
                   -\frac{\displaystyle G\Msat}{\displaystyle \rt}\left[1 + \frac{\left(1+\xt\right)^2}{\xt}\left(\frac{1}{1+x}-\frac{1}{1+\xt}\right)\right], & x\leq \xt.\\
                   \null &\null \\
                   -\frac{\displaystyle G\Msat}{\displaystyle r}, & x> \xt.
                   \end{cases} \\
        \rho(x) &= \begin{cases}
                   \frac{\displaystyle A}{\displaystyle x(1+x)^3}, & x\leq \xt.\\
                   \null&\null\\
                   0, & x > \xt.
                   \end{cases} \\
    \end{aligned}
\end{equation}
The density normalisation $A$ is related to the total satellite mass $\Msat$ by
\begin{equation}
    A = \frac{(1+x_t)^2}{x_t^2} \frac{\Msat}{2\pi a^3},
\end{equation}
The mass enclosed within a reduced radius $x$ is then
\begin{equation}
\label{E:MassFrac}
    m(x)=\begin{cases}
        \Msat\frac{\displaystyle x^2(1+x_t)^2}{\displaystyle x_t^2(1+x)^2}, & \text{if $x\leq x_t$}.\\
        \null &\null\\
        \Msat, & \text{otherwise}.
    \end{cases}
\end{equation}
For all satellites, we adopt truncation radii of $\rt=10\aS$, or equivalently $\xt=10$.

The acceleration on any given test particle due to the satellite can then be calculated from the above relations. For self-consistency, the initial phase-space distribution of the tracer particles is that of a truncated Hernquist profile (see Section \ref{S:Methods:Tracers} for further details). Of course, this self-consistency is lost as the simulation advances in time, as many of the tracer particles are tidally removed by the Milky Way, but our assumed satellite potential remains unchanged in mass and shape. However, we will show in \S\ref{S:CodeValidation} that this assumption of an unchanging satellite potential is largely harmless.

\subsection{Fifth Forces}
\label{S:Models:FifthForce}

In addition to gravity, the satellite and the tracer particles also experience accelerations due to the fifth force. The satellite feels a fifth force sourced by the Milky Way, while the tracer particles also feel a fifth force sourced by the satellite. We assume spherical fifth force profiles in both cases. For the satellite, this is consistent with its gravitational potential, although the sphericity of the satellite may be distorted by its tidal disruption. For the Milky Way, the spherical symmetry is inconsistent with the presence of the disc. The scalar field profiles of disc galaxies have correspondingly discoid shapes \citep{Naik2018}. However, the scalar field profile is roughly spherical when $\rscrMW$ is much larger than the disc scale radius of 6.5 kpc. In particular, using the $f(R)$ scalar field solver described in \S\ref{S:Results:Future}, we find that fifth force profiles in the Milky Way (assuming a spherical dark matter halo) only become appreciably aspherical for $\log_{10}|f_{R0}| \gtrsim -6.2$, and so the spherical approximation is robust in the parameter regimes we mostly focus on in this article.

Eq.~(\ref{E:FifthForceExtendedParticle}) can be rewritten to give the expression for the modified gravity acceleration due to the satellite on tracer particle $i$, situated at position $\bm{x}$,
\begin{equation}
    \bm{a}^i_{5, \mathrm{sat}}(\bm{x}) = 2\beta^2 Q_i Q_\mathrm{sat}(r) \bm{a}_{\mathrm{N, sat}}(\bm{x}),
\end{equation}
where $\beta$ is the coupling strength of the fifth force (an input parameter of our simulations), $\bm{a}_{\mathrm{N, sat}}$ is the Newtonian acceleration due to the satellite, and $Q_i$ and $Q_\mathrm{sat}(r)$ are the scalar charges of particle $i$ and the satellite respectively. The latter is given by
\begin{equation}
\label{E:Qsat}
    Q_\mathrm{sat}(r) = \begin{cases}
        1-\frac{\displaystyle m(\rscrsat)}{\displaystyle \Msat}, & \text{if $r\geq \rt$}.\\
        \null &\null\\
        1-\frac{\displaystyle m(\rscrsat)}{\displaystyle m(r)}, & \text{if $\rt > r\geq\rscrsat$}.\\
        \null &\null\\
        0, & \text{otherwise}.
    \end{cases}
\end{equation}
Here, $m(r)$ is the satellite mass enclosed by radius $r$, and $\rscrsat$ is its screening radius. $Q_i$, meanwhile, differs between the particle types. As we assume the stars are fully screened against the fifth force, $Q_i=0$ for the star tracer particles. On the other hand, we take $Q_i=1$ for the dark matter tracer particles, which we assume to be a diffuse, unscreened component.

Similarly, the modified gravity acceleration due to the Milky Way on particle $i$ (which can now also represent the satellite) at $\bm{x}$ is given by
\begin{equation}
\label{E:aMW}
    \bm{a}^i_{5, \mathrm{MW}}(\bm{x}) = 2\beta^2 Q_i Q_\mathrm{MW}(r) \bm{a}_{\mathrm{N, MW}}(\bm{x}),
\end{equation}
where the symbols have analogous meanings to those above. The scalar charge of the Milky Way is given by
\begin{equation}
    Q_\mathrm{MW}(r) = \begin{cases}
        1-\frac{\displaystyle M(\rscrMW)}{\displaystyle M(r)}, & \text{if $r\geq\rscrMW$}. \\
        \null &\null \\
        0, & \text{otherwise}.
    \end{cases}
\end{equation}
If particle $i$ represents the satellite, then we take the limiting value of the satellite scalar charge $Q_i=Q_\mathrm{sat}(r=\rt)$. This is valid as long as the the Milky Way centre does not fall within the truncation radius of the satellite centre, which does not happen in any of our simulations.

The formalism given in this subsection demonstrates the utility of truncating the mass profile of the satellite. By so doing, we have made it straightforward to model the satellite as being fully screened ($\rscrsat=r_t$), fully unscreened ($\rscrsat=0$), or partially screened ($0 < \rscrsat < r_t$).

It is worth remarking that we have used the superposition principle to compute the joint fifth force of Milky Way and satellite on the tracer particles. Strictly speaking, the superposition is not valid in highly non-linear theories of gravity like chameleon gravity. In particular, environmental screening can affect the screening radii of objects. Linearity is, however, restored once the screening radii are fixed (as we do by hand), so that from that point on we can apply the superposition principle for computing the joint fifth force.

\begin{figure}[htp]
    \centering
    \includegraphics{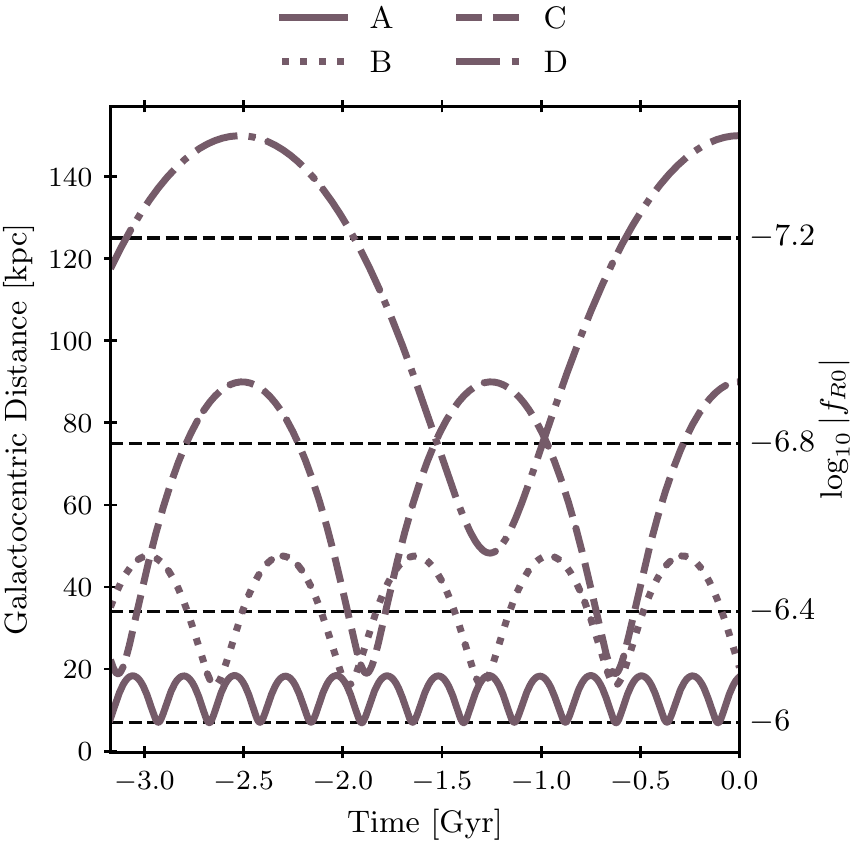}
    \caption{For the 4 satellites described in Table \ref{T:StreamICs}, we show the orbital evolution over $10^{17}$ seconds ($\sim 3$ Gyr) under standard gravity. The horizontal dashed lines indicate the Milky Way screening radii under Hu-Sawicki $f(R)$ gravity for various different values of the theory parameter $f_{R0}$ (the values of $\log_{10}|f_{R0}|$ are shown at the right hand side of the panel  (see Section~\ref{S:Methods:Simulations} for details about the calculation of these screening radii). This figure illustrates the range of distances probed by tidal streams, and gives an idea of the possible constraints achievable for chameleon gravity theories.}
    \label{F:OrbitDistances}
\end{figure}

\begin{table}[htp]
    \caption{Parameters for each of the 4 progenitors. Here, $\bm{x}_0$ and $\bm{v}_0$ give the present position and velocity respectively (note that we run the simulations backwards then forwards again, so that the satellites end at $\bm{x}_0$ and $\bm{v}_0$), $a$ and $\Msat$ are the Hernquist scale radius and total mass of the satellite, and $t_\mathrm{max}$ is the total time over which each simulation is run; the farther orbits require more time to undergo an appreciable number of orbital periods. Note that $10^{17}$ seconds is $\sim 3$ Gyr. The parameters for Satellite A resemble the Pal-5 stream, B the Sagittarius stream, C the Orphan stream, and D a hypothetical stream at large distance.}
    \begin{ruledtabular}
    \begin{tabular}{l l l l l l l}
    ID  & $\bm{x}_0$         & $\bm{v}_0$       & $a$   & $\Msat$           & $t_\mathrm{max}$  \\
        & (kpc)              & (km/s)           & (kpc) & ($10^8 M_\odot$)  & ($10^{17}$ s)     \\ \hline
    A   & (7.7, 0.2, 16.4)   & (-44, -117, -16) & 0.01  & 0.0003            & 1                 \\
    B   & (19.0, 2.7, -6.9)  & (230, -35, 195)  & 0.5   & 5                 & 1                 \\
    C   & (90, 0, 0)         & (0, 0, 80)       & 0.5   & 2.5               & 1.5               \\
    D   & (150, 0, 0)        & (0, 0 , 100)     & 1     & 5                 & 2.5               \\
    \end{tabular}
    \end{ruledtabular}
    \label{T:StreamICs}
\end{table}

\section{Methods}
\label{S:Methods}

Approximate methods for quickly generating realistic streams by stripping stars at the tidal radius of a progenitor are now well established \citep{Lane2012,Kupper2012,Gibbons2014}. The methods work as restricted N-body simulations, in which we follow the orbital evolution of a large number of massless tracer particles. The stream particles are integrated in a fixed Galactic potential, together with the potential of the moving satellite. This method robustly reproduces the morphology of streams, in particular the locations of the apocentres of the leading and trailing branches, yet provides two to three orders of magnitude speed-up compared to conventional N-body experiments \citep{Gibbons2014}. The main extension of our code here is that it incorporates an optional fifth force due to the chameleon field.

All of our code is made publicly available as the \texttt{python 3} package \texttt{smoggy} (Streams under MOdified GravitY) \footnote{\url{https://github.com/aneeshnaik/smoggy}}. Animations of the simulations depicted in Figures \ref{F:StandardGravityImages}, \ref{F:TimeSeries}, \ref{F:InterestingSignatures}, and \ref{F:StreamCDF} are given as Supplemental Material accompanying this article \footnote{See Supplemental Material at \textbf{[URL will be inserted by publisher]} for animations of the simulations depicted in Figures \ref{F:StandardGravityImages}, \ref{F:TimeSeries}, \ref{F:InterestingSignatures}, and \ref{F:StreamCDF}}.

\subsection{Tracer Particles}
\label{S:Methods:Tracers}

To generate the initial phase space distribution of N tracer particles, we use a Markov Chain Monte Carlo technique to generate $2N$ samples from possible equilibrium distribution functions (DFs) for the Hernquist model. The choice of equilibrium includes the isotropic DF \citep{Hernquist1990}
\begin{equation}
\begin{split}
    & f(\tE) = \frac{1}{\sqrt{2}\left(2\pi\right)^3\left(GM'\aS\right)^{3/2}}\frac{\sqrt{\tE}}{\left(1-\tE\right)^2} \\
    & \times \left[\left(1-2\tE\right)\left(8\tE^2-8\tE-3\right)+\frac{3\sin^{-1}\sqrt{\tE}}{\sqrt{\tE\left(1-\tE\right)}}\right],
\end{split}
\end{equation}
and the radially anisotropic DF~\citep{Evans2006}
\begin{equation}
    f(\tE) = \frac{3}{4\pi^3\aS}\frac{\tE}{GL}.
\end{equation}
Here, $E$ is the specific (binding) energy of a particle, $\aS$ is the scale radius, $M' = (1+x_t)^2\Msat/x_t^2$ is the untruncated mass of the satellite, while $\tE = E\aS/GM'$ is the dimensionless binding energy. The DFs differ in the anisotropy of the velocity distributions. In fact, our simulations show similar results for stream generation, irrespective of the anisotropy, so the choice of equilibrium is not so important.

Given these $2N$ samples, we integrate the orbits of the particles in the satellite potential (i.e. neglecting fifth forces and the Milky Way) for $10^{17}$ seconds ($\approx 3$ Gyr). At the end of this relaxation phase, we randomly downsample $N$ of these particles, excluding any particles for which the orbit ever strayed beyond the truncation radius. This gives a suitable equilibrium distribution of positions and velocities for the test particles in our simulations.

All of our simulations incorporate 10000 DM particles and 10000 star particles. This equality does not encode any assumptions about the underlying stellar/DM mass fraction of the satellite; the particles are merely massless samples of the distribution, and the satellite is assumed to be dark matter dominated.

This procedure has omitted the fifth force altogether. This is appropriate for the star particles which, by assumption, do not experience the fifth force. For the dark matter however, it is less self-consistent. We have experimented with including a fifth force, both in the distribution function and in the relaxation phase described in the previous paragraph. This leads overall to $\sim 10 \%$ increases in the number of dark matter particles being stripped from the progenitor during the main simulation, but no appreciable morphological change to the dark matter streams.

Note also that this procedure makes the unrealistic assumption that the stars and dark matter have the same spatial distribution. However, we have experimented with drawing the stars from more compact initial distributions than the dark matter, and found no appreciable difference in our results.

\subsection{Orbit Integration}

To calculate the trajectories of the various particles, we use a second-order leapfrog integrator. Under such a scheme, the velocities $\bm{v}$ and positions $\bm{x}$ of the particles are updated at each timestep $i$ via
\begin{equation}
\label{E:LeapfrogUpdate}
    \begin{aligned}
        \bm{v}_{i+1/2} &= \bm{v}_{i-1/2} + \bm{a}(\bm{x}_i)\Delta t,\\
        \bm{x}_{i+1} &= \bm{x}_i + \bm{v}_{i+1/2}\Delta t,
    \end{aligned}
\end{equation}
where $\Delta t$ represents the timestep size, and $\bm{a}(\bm{x})$ represents the accelerations calculated using the expressions given in Sections \ref{S:Models:MW}, \ref{S:Models:Satellite}, and \ref{S:Models:FifthForce}. At the start of the simulation (i.e. timestep $i=0$), the `desynchronised' velocities $\bm{v}_{-1/2}$ are obtained using
\begin{equation}
    \bm{v}_{-1/2} = \bm{v}_0 - \frac{1}{2}\bm{a}(\bm{x}_0)\Delta t.
\end{equation}
From here, Eq.~(\ref{E:LeapfrogUpdate}) can be used repeatedly to advance the system in time.

Our method for choosing the timestep size $\Delta t$ is as follows. We calculate the total energies of all particles at the start and end of the relaxation phase described in Section \ref{S:Methods:Tracers}, in which the orbits are integrated in the satellite potential for $10^{17}$ seconds. We repeat the relaxation phase, iteratively reducing the timestep size, until the energies of all particles are conserved to within 2\%. Through experimentation, we found that energy conservation is a good proxy for numerical convergence and this 2\% criterion gives accurate, converged results. With this criterion, we find that angular momentum is conserved to an even greater precision, with a maximum fractional deviation of ${\sim}10^{-4}$. The final timestep size chosen by this process is then used again for the main simulation. In practice, we find $\Delta t \sim \mathcal{O}(10^{11})$ seconds typically.

\subsection{Simulations}
\label{S:Methods:Simulations}

We simulate the generation of streams from 4 progenitors. Satellite A is inspired by the Palomar 5 stream \cite{Pearson2017}, B the Sagittarius stream \cite{Law2010}, C the Orphan stream \cite{Koposov2019}, and D is a hypothetical stream at large Galactocentric distance, of the kind that is likely to be found in the later Gaia data releases. The parameters for these 4 progenitors are given in Table \ref{T:StreamICs}.

Figure \ref{F:OrbitDistances} shows the evolution of the orbits over $\sim 3$ Gyr for each of the 4 satellites, under standard gravity. Also shown are lines indicating the disc-plane Milky Way screening radii for a range of values of $f_{R0}$. These calculations were performed using the scalar field solver within the $f(R)$ N-body code \textsc{mg-gadget} \citep{Puchwein2013} for the Milky Way model described in \ref{S:Models:MW}. We demonstrate later that significant stream asymmetries develop when the orbit is mostly outside the Milky Way screening radius, so these lines give a preview of the modified gravity constraints achievable.

For each satellite, we explore a variety of modified gravity scenarios by varying 3 input parameters: the coupling strength $\beta$, the satellite screening radius $\rscrsat$, and the Milky Way screening radius $\rscrMW$. First, we consider 4 coupling strengths: $\beta=\{0.1, 0.2, 0.3, 0.4\}$. The strength of the fifth force relative to gravity is given by $2\beta^2$, so this corresponds to the range from $2\% - 32\%$. The most extreme case can therefore be used as an approximate analogue for $f(R)$ gravity, where the strength of the fifth force is $1/3$ that of gravity.

For the satellite screening radius, we explore a range of regimes, from fully screened to fully unscreened, and encompassing a variety of partially screened regimes in between. Using the upper case of Eq.~(\ref{E:Qsat}), we recast the screening radius $\rscrsat$ as the scalar charge $Q_\mathrm{sat}$, and consider a range of values of $\Qsat$ from 0 to 1 in steps of 0.1. We recall that $\Qsat=0$ corresponds to the fully screened case, so here $\rscrsat=10a$, where $a$ is the Hernquist scale radius of the satellite in question. $\Qsat=1$ is the fully unscreened case, so $\rscrsat=0$.

Finally, we consider a range of values for the Milky Way screening radius $\rscrMW$. As the orbital distances of each satellite are different, it is useful to select a different range of values for $\rscrMW$ for each satellite. For each satellite, we define a maximum screening radius $\rscrmax$, approximately equal to the apocentric distance of the orbit under standard gravity. These values are $\rscrmax=20, 50, 90, 150$ kpc for satellites A, B, C, and D respectively. Then, we choose a range of 11 values such that $\rscrMW/ \rscrmax$ runs from 0 to 1 in steps of 0.1.

Altogether, we run 485 simulations for each satellite: $4\times 11 \times 11 = 484$ modified gravity simulations plus one standard gravity ($\beta=0$) simulation.

\subsection{Assumptions}
\label{S:Methods:Assumptions}

The previous subsections have given details about the various parts of our code, but for clarity we provide a list of all of our simplifying assumptions:

\begin{figure*}[htp]
    \centering
    \includegraphics{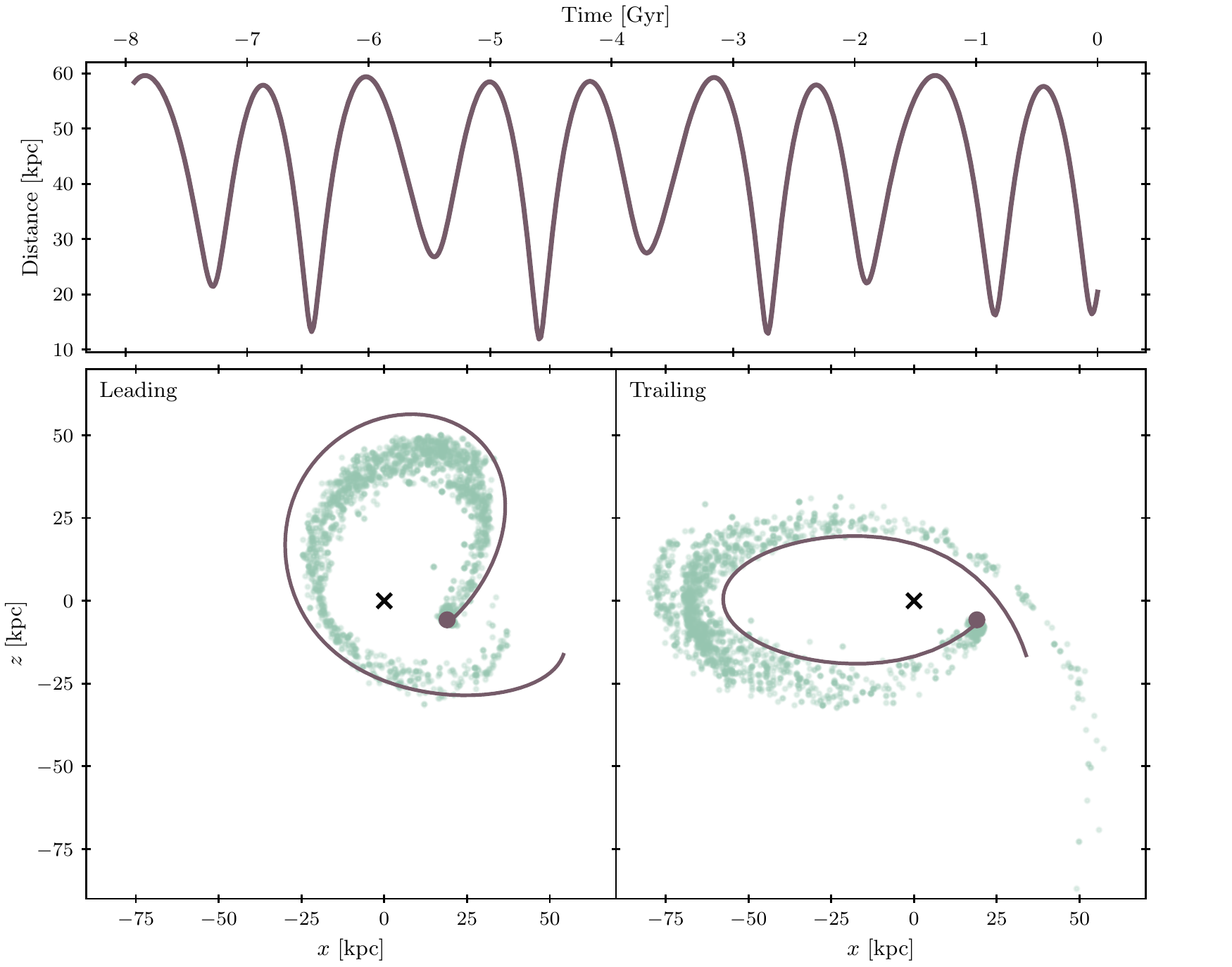}
    \caption{Our reproduction of a simulation from \citet{Law2010} \textit{Top:} Distance of the simulated Sagittarius dwarf from the Galactic centre over 8 Gyr (to be compared to the results in Figure 7 from Ref.~\cite{Law2010}). \textit{Bottom left and right:} First wrap of the leading and trailing streams respectively (to be compared to the results in the two left-hand panels of Figure 8 of Ref.~\cite{Law2010}). The curve represents the orbital path of the satellite, culminating in the current position of the Sagittarius dwarf, represented by the filled circle. The green points are the positions of the simulation particles. The satellite orbit has been integrated over 3 Gyr up to the present day, so the morphology of the streams should resemble only the orange and magenta particles from the original figure. This successful reproduction of literature results serves as a test of our code, and checks several of our simplifying assumptions.}
    \label{F:LM10}
\end{figure*}

\begin{enumerate}

    \item\label{A:SelfGravity} We neglect self-gravity between the tracer particles, both before and after they are stripped from the satellite, as is typical in Lagrange stripping codes \citep{Gibbons2014,Bowden2015}.

    \item\label{A:Hernquist} We assume the gravitational attraction on the tracer particles due to the satellite can be approximated as that due to a (truncated) Hernquist sphere, whose orbit is only governed by the Milky Way potential. This assumption has been verified against full N-body simulations of stream formation by others \citep{Lane2012,Kupper2012,Gibbons2014}.

    \item\label{A:ConstantSatPotential} We assume the depth and radial extent of the satellite potential well does not change over time. While this assumption could be relaxed in the standard gravity case, it is a greatly helpful one in the chameleon case. Thus, to allow a fair comparison between results in the two cases, we make the assumption universally.

    \item\label{A:MWModel} We assume a static, axisymmetric model for the Milky Way potential, composed of a disc, bulge, and halo.  Dynamical friction is therefore not modelled, though the effect is negligible at these low mass ratios \citep{Boylan2008}. We neglect any effects due to the Large Magellanic Cloud or other Milky Way satellites (cf. \cite{Koposov2019})

    \item\label{A:DMDominated} While we typically sample equal numbers of stellar and dark matter particles, we assume the mass profiles of our satellites to be dark matter dominated. So, the satellites feel the full fifth force in the absence of screening.

    \item\label{A:P1P2Profiles} The initial density profile and kinematics of the stellar and dark matter particles in the satellites are assumed to be the same. This simplifies the fifth force calculation, and allows us to ensure any difference in the stellar and dark matter streams is due to the fifth force rather than initial conditions. As described in Section \ref{S:Methods:Tracers}, we have experimented with sampling the star particles from radially more compact distributions than the dark matter, and found no significant difference in results.

    \setcounter{AssumptionListCounter}{\value{enumi}}
\end{enumerate}

Assumptions (\ref{A:SelfGravity})-(\ref{A:P1P2Profiles}) apply equally in the standard gravity and modified gravity simulations. The following three assumptions, however, apply only in the simulations including a fifth force.

\begin{enumerate}
    \setcounter{enumi}{\value{AssumptionListCounter}}
    \item We adopt spherical fifth force profiles around both the Milky Way and the satellite, despite the Milky Way potential being non-spherical. As discussed in \S\ref{S:Models:FifthForce}, this is valid when the the MW screening radius is larger than the Galactic disc, $\log_{10}|f_{R0}| \lesssim -6.2$.

    \item Furthermore, we assume this spherical screening surface of the satellite remains fixed throughout the satellite's orbit. In reality, the radius would vary as the Galactocentric distance of the satellite changes, due to environmental screening, and the shape of the screening surface (and surrounding fifth force profile) would likely become aspherical as the satellite approached the Milky Way's screening radius and non-linear effects warp the screening surface.

    \item The Compton wavelength of the scalar field is assumed to be much larger than relevant length scales. In the context of Hu-Sawicki $f(R)$ gravity, the Compton wavelength is given by $\lambda_\mathrm{C} \approx 32 \sqrt{|f_{R0}|/10^{-4}}$ Mpc \citep{Cabre2012}, so this assumption starts to break down at around $f_{R0} \sim 10^{-8}.$

\end{enumerate}

\begin{figure*}[htp]
    \centering
    \includegraphics{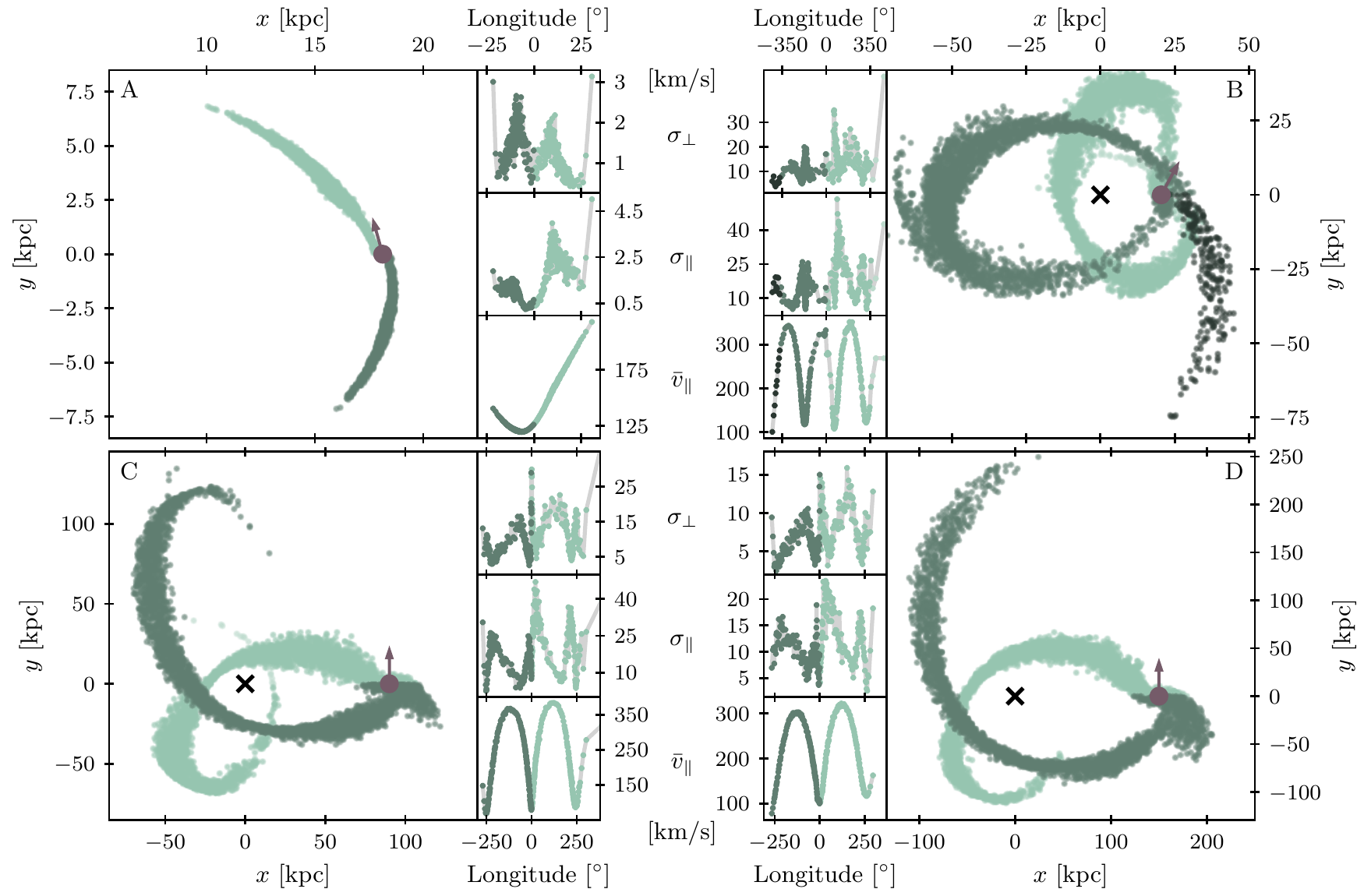}
    \caption{The simulated streams under standard gravity. The four quarters represent our 4 satellites: A (\textit{upper left}), B (\textit{upper right}), C (\textit{lower left}), and D (\textit{lower right}). In each quarter, the largest subpanel shows an image of all stream particles in the orbital plane, at the end of the simulation. No distinction is made between star and dark matter particles. The colours differentiate leading and trailing streams, with the darker shade being the trailing stream. For Satellite B, additional shades are used to distinguish multiple wraps. The black cross shows the position of the centre of the Milky Way, while the filled circle shows the final position of the Satellite, with an arrow indicating its instantaneous direction of travel. The side-panels show three quantities calculated in bins of particles: average velocity along the stream, velocity dispersion along the stream, and velocity dispersion perpendicular to the stream. Here again, the colours differentiate leading and trailing streams. In every case, the orbital plane is defined such that the Satellite is on the $x$-axis, moving in the positive $y$-direction. Animations of the 4 simulations depicted in this figure are included in the Supplemental Material accompanying this article.}
    \label{F:StandardGravityImages}
\end{figure*}

\section{Code Validation}
\label{S:CodeValidation}

As validation, we compare the results of our code for disruption of the Sagittarius dwarf galaxy under standard Newtonian gravity with the results of \citet{Law2010}. They simulate the formation of the stream using a full N-body disintegration of the satellite in a static Milky Way potential, so assumptions (1)-(3) in the list in \S\ref{S:Methods} are not made in their work. In other words, the gravitational attractions of the satellite and stream are there treated in fully self-consistent manner.

To set up this test, we adopt the Milky Way potential of Ref.~\cite{Law2010}, i.e. a Hernquist bulge, a Miyamoto-Nagai disc, and a triaxial logarithmic dark matter halo. The parameters and initial conditions for the satellite are the same as those for Satellite B, given in Table \ref{T:StreamICs}.

As a first test, we integrate the orbit of the satellite in this potential backwards for $2.5 \times 10^{17}$ seconds ($\sim 8$ Gyr). The distance of the satellite from the Galactic centre as a function of time is shown in the upper panel of Figure \ref{F:LM10}. This shows excellent agreement with Figure 7 from \citet{Law2010}.

It is also desirable to check the morphology of the streams generated with our method. As a second test, we integrate the orbit of the satellite backwards for $10^{17}$ seconds ($\sim 3$ Gyr), and then forwards again with 16000 tracer particles. The resulting leading and trailing streams from this simulation are shown in the lower pair of panels in Figure \ref{F:LM10}. The detailed morphologies of these streams closely resemble those of the streams depicted in Figure 8 of \citet{Law2010}, considering only the orange and magenta particles in that figure (i.e., particles liberated within the last 3 Gyr).

Despite this reassuring agreement between the results from our simplified code and those from full N-body simulations, it is worth noting that several of the assumptions stated in \S\ref{S:Methods:Assumptions} are not addressed by this test. In particular, this test does not validate the assumptions made in the treatment of the fifth force. However, the aim of the present work is to provide a qualitative understanding of the effects of chameleon gravity on stellar streams. Future work aiming to derive quantitative constraints from observational data will likely require either a relaxation or a more careful justification of some of those assumptions.

\begin{figure}[htp]
    \centering
    \includegraphics{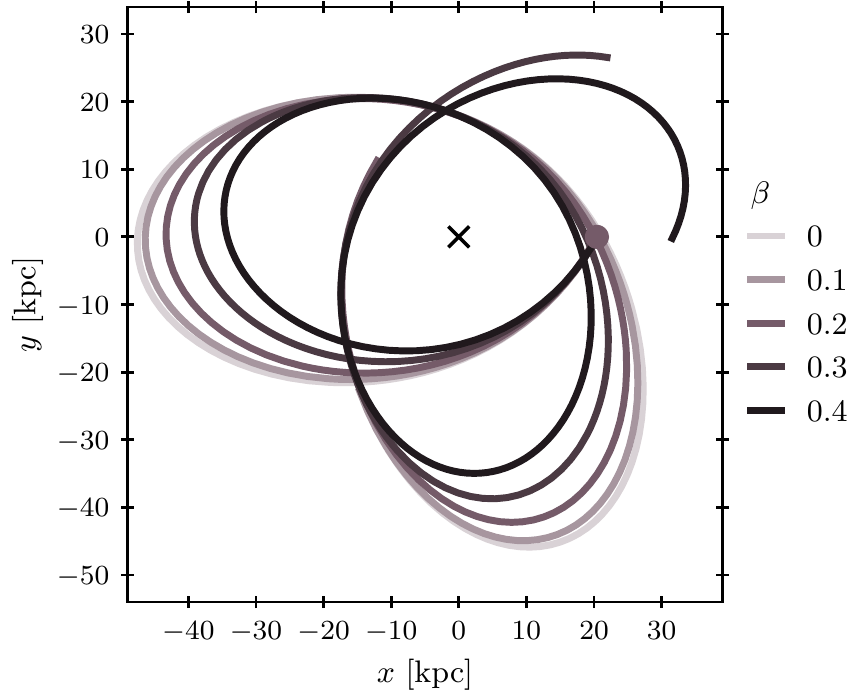}
    \caption{Satellite B's orbit in its orbital plane, shown for a range of $\beta$ with $\rscrsat=\rscrMW=0$. The cross indicates the Galactic centre and the filled circle shows the final position of the satellite, i.e. the current observed position of the Sagittarius dwarf galaxy. This figure illustrates the effect of an unscreened fifth force on orbital shapes for a fixed final position.}
    \label{F:OrbitalShapesBeta}
\end{figure}

\begin{figure*}[htp]
    \centering
    \includegraphics{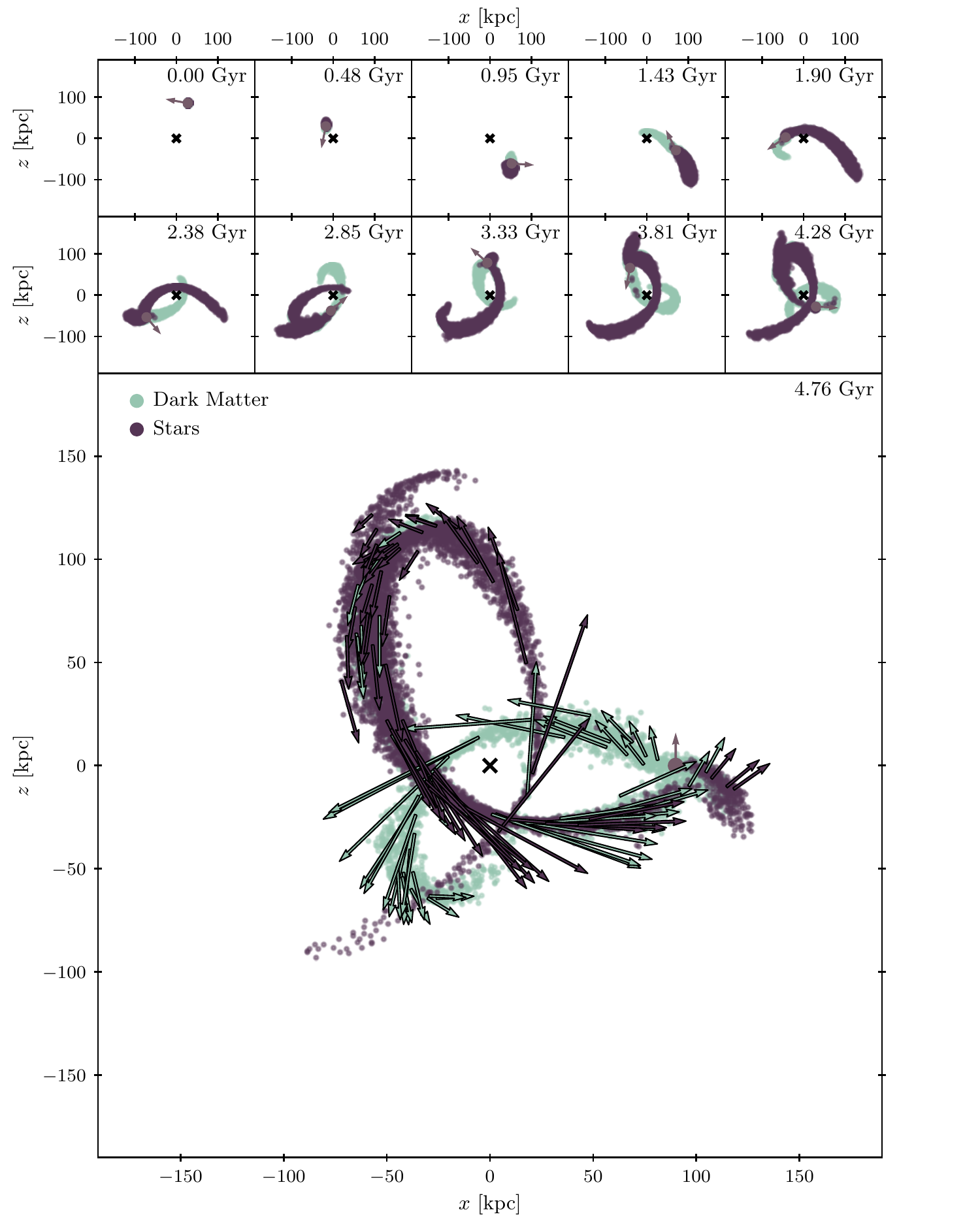}
    \caption{The simulation depicted here is Satellite C with no screening and a fifth force coupling only to dark matter with $\beta=0.2$. The large panel shows an image of the stellar (purple) and dark matter (green) streams at the end of the simulation, while the smaller panels above show the evolution over time. The interval between images is $1.5 \times 10^{16}$ seconds ($\sim 0.48$ Gyr, as labelled). The cross and large filled circle respectively indicate the positions of the Milky Way and satellite centres. In the large panel, 50 unbound particles have been randomly chosen from each species, and arrows of the corresponding colour are shown indicating their velocities. An animation of this simulation is included in the Supplemental Material accompanying this article. This figure shows the formation of an asymmetric stellar stream over time.}
    \label{F:TimeSeries}
\end{figure*}

\begin{figure*}[htp]
    \centering
    \includegraphics{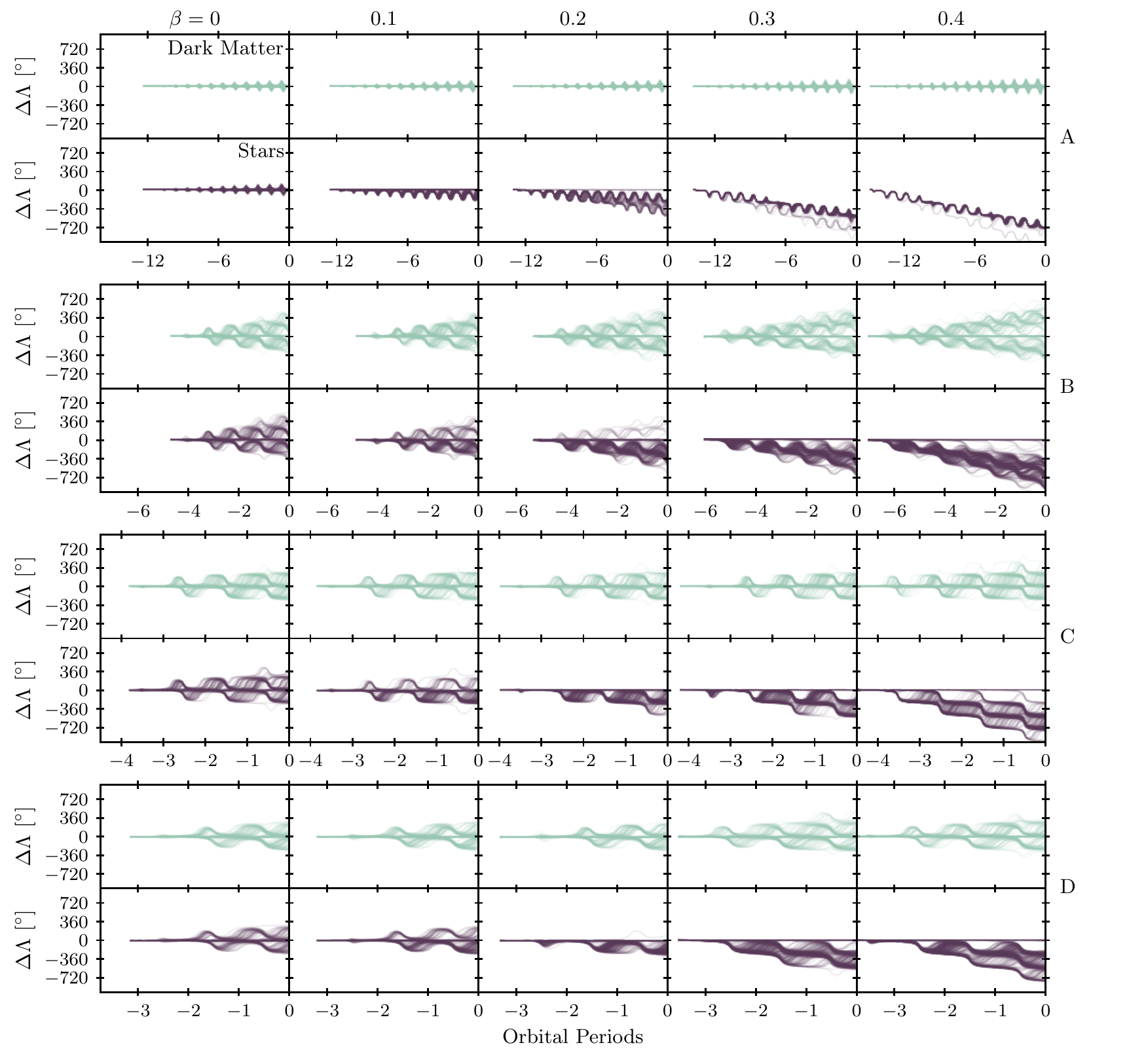}
    \caption{The longitude difference $\Delta\Lambda=\Lambda - \Lambda_\mathrm{sat}$ as a function of time for all 4 satellites without screening. Each column shows a
    different fifth force coupling from $\beta=0$ to 0.4 in steps of 0.1. Here, $\Lambda$ is longitude in the orbital plane of the satellite, increasing in the direction of the satellite's orbit. Lines are drawn for 500 star and 500 DM particles in each simulation, i.e. 1 in 200 particles are randomly sampled. Complementing Figure \ref{F:TimeSeries}, this figure shows the development over time of the asymmetry of the stellar streams, and the increased magnitude of this effect with $\beta$.}
    \label{F:UnscreenedLongitudes}
\end{figure*}

\begin{figure}[htp]
    \centering
    \includegraphics{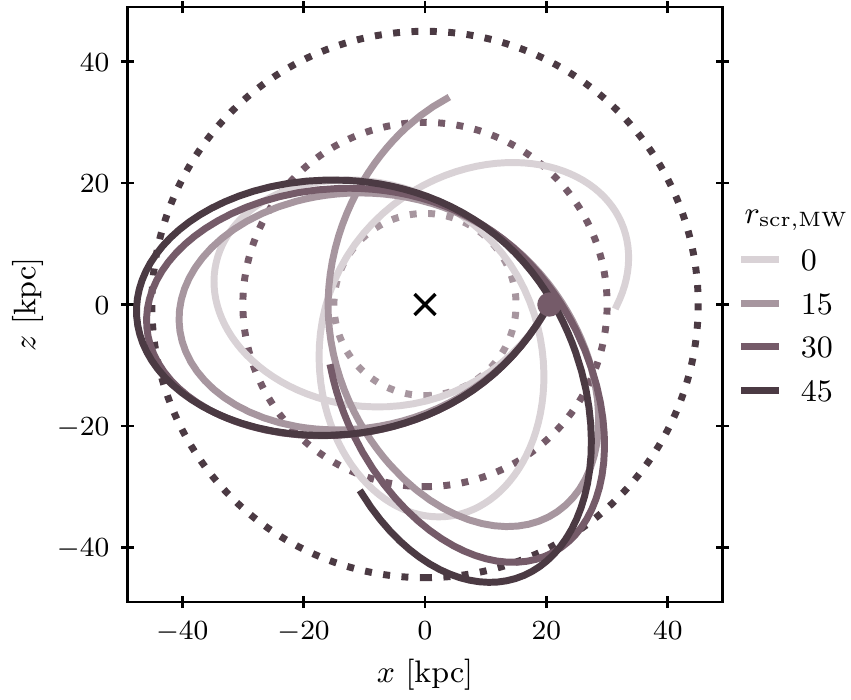}
    \caption{Satellite B's orbit in its orbital plane, shown for $\rscrMW=0, 8, 24, 50, 85$ kpc, and $\rscrsat=0, \beta=0.4$. The dotted circles indicate the position of the screening radius in each case. The cross indicates the Galactic centre and the filled circle shows the final position of the satellite, i.e. the current observed position of the Sagittarius dwarf galaxy. This figure illustrates the effect of a Milky Way screening radius on the satellite orbital shapes.}
    \label{F:OrbitalShapesRscrMW}
\end{figure}

\begin{figure}[htp]
    \centering
    \includegraphics{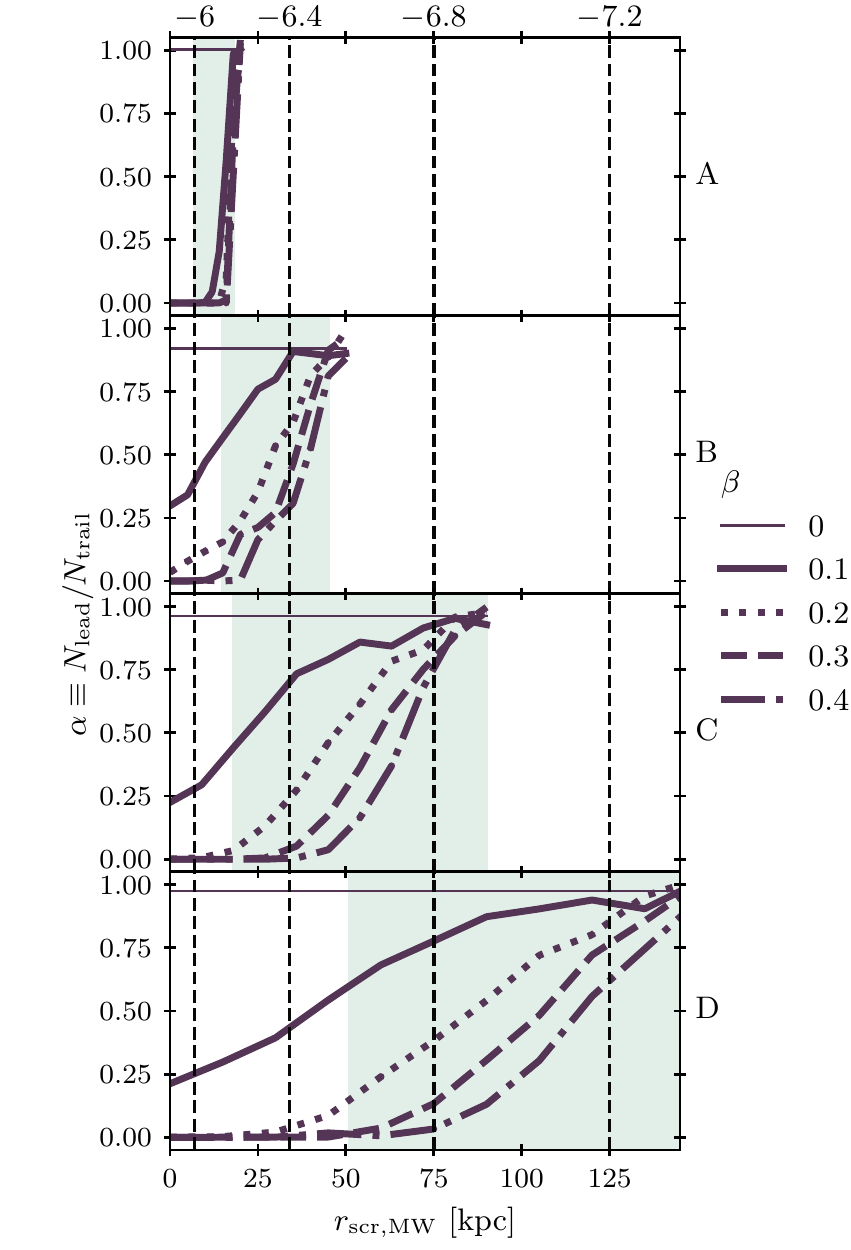}
    \caption{The asymmetry parameter $\alpha \equiv N_\mathrm{lead}/N_\mathrm{trail}$, for all simulations with $\Qsat=1$. The 4 panels correspond to the 4 satellites and the different textures of line correspond to different values of $\beta$. In each panel, the shaded region indicates the radial range of the satellite's orbit. As with the horizontal lines in Figure \ref{F:OrbitDistances}, the vertical dashed lines here show the locations of Milky Way screening radii for various values of $\log_{10}{|f_{R0}|}$. This figure shows the Milky Way screening radius can affect the stream asymmetry. Streams at larger Galactocentric distances are sensitive to larger screening radii, and therefore weaker modified gravity regimes.}
    \label{F:RscrMWComparison}
\end{figure}
\begin{figure}[htp]
    \centering
    \includegraphics{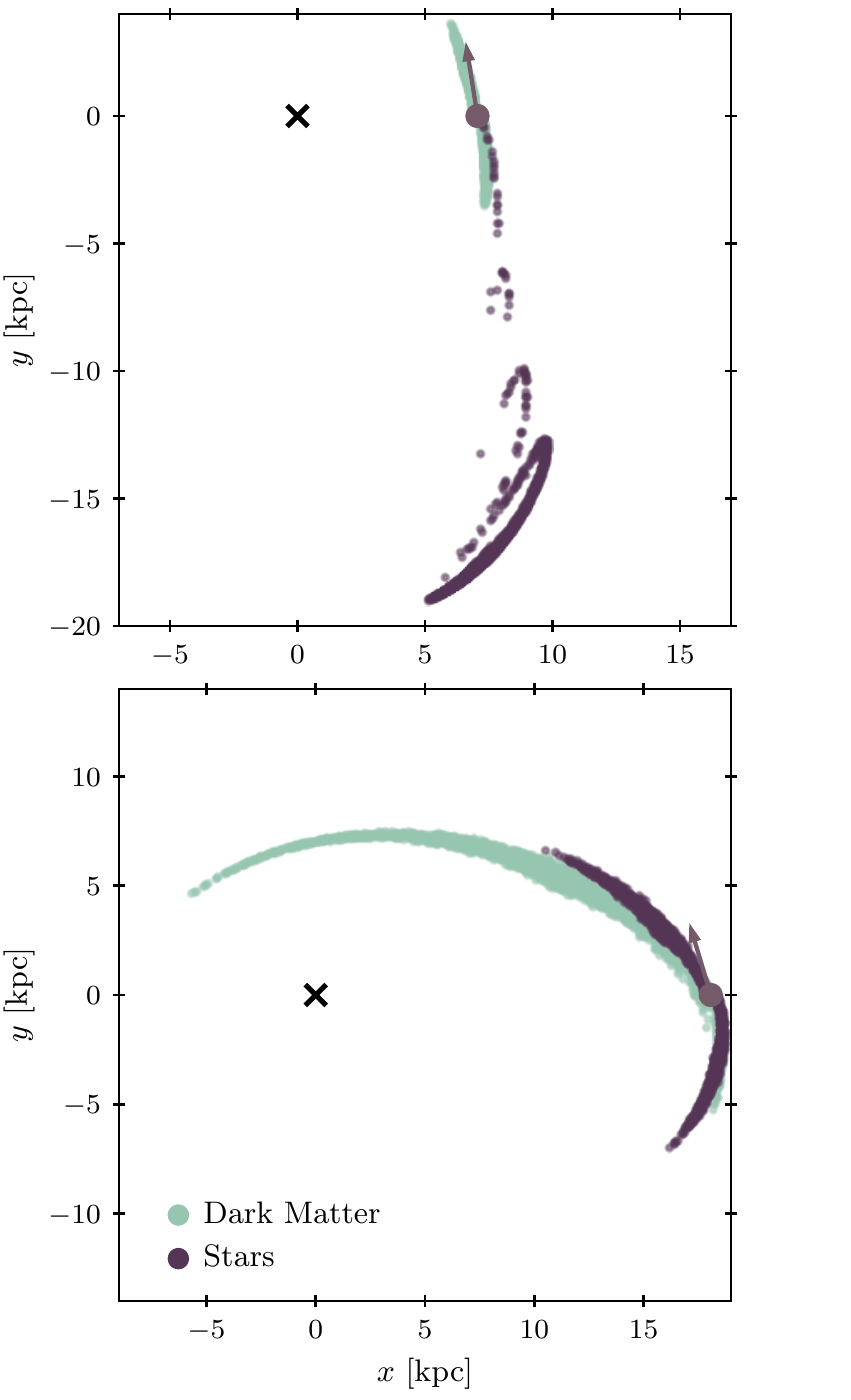}
    \caption{\textit{Top:} An image from a simulation of Satellite A, $\beta=0.4$, $\Qsat=1$, and $\rscrMW=10$ kpc. \textit{Bottom:} An image from another simulation of Satellite A, $\beta=0.1$, $\Qsat=0$, and $\rscrMW=4$ kpc. Animations of the 2 simulations depicted in this figure are included in the Supplemental Material accompanying this article. This figure shows some interesting signatures of screened modified gravity other than the stellar asymmetry we have discussed in previous figures.}
    \label{F:InterestingSignatures}
\end{figure}
\begin{figure*}[htp]
    \centering
    \includegraphics{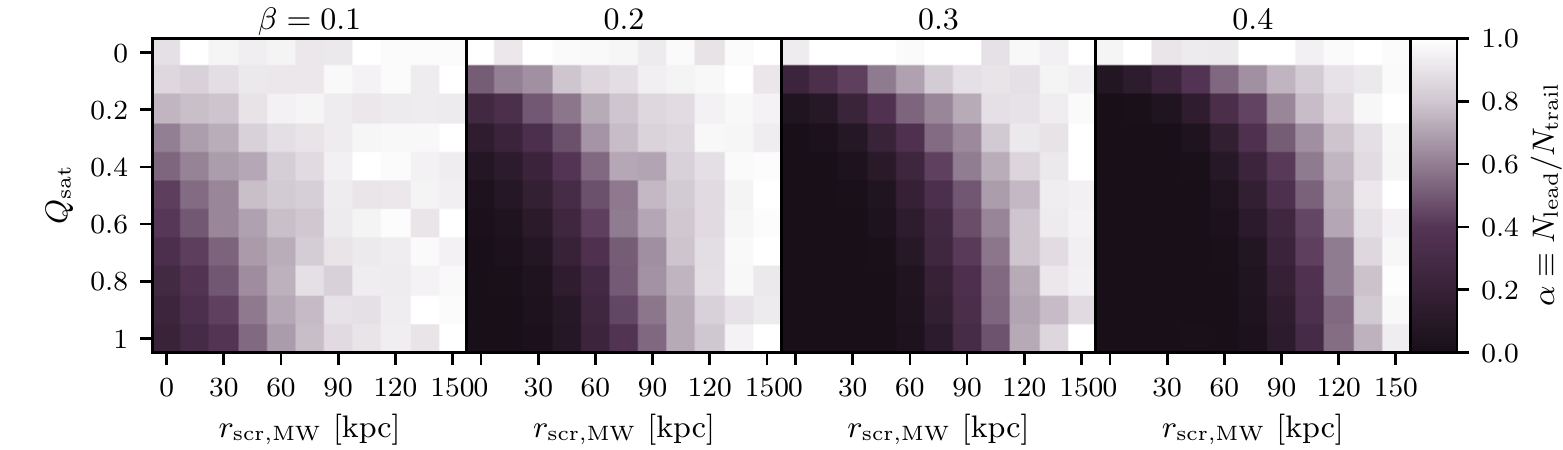}
    \caption{The asymmetry parameter $\alpha \equiv N_\mathrm{lead}/N_\mathrm{trail}$ for the unbound stellar particles in all simulations of satellite D with screening, shown here as a function of $\Qsat$ and $\rscrMW$, with different panels corresponding to different values of $\beta$. This figure shows the effects of varying all of our parameters on the stream asymmetries.}
    \label{F:DAsymmetry}
\end{figure*}

\begin{figure}[htp]
    \centering
    \includegraphics{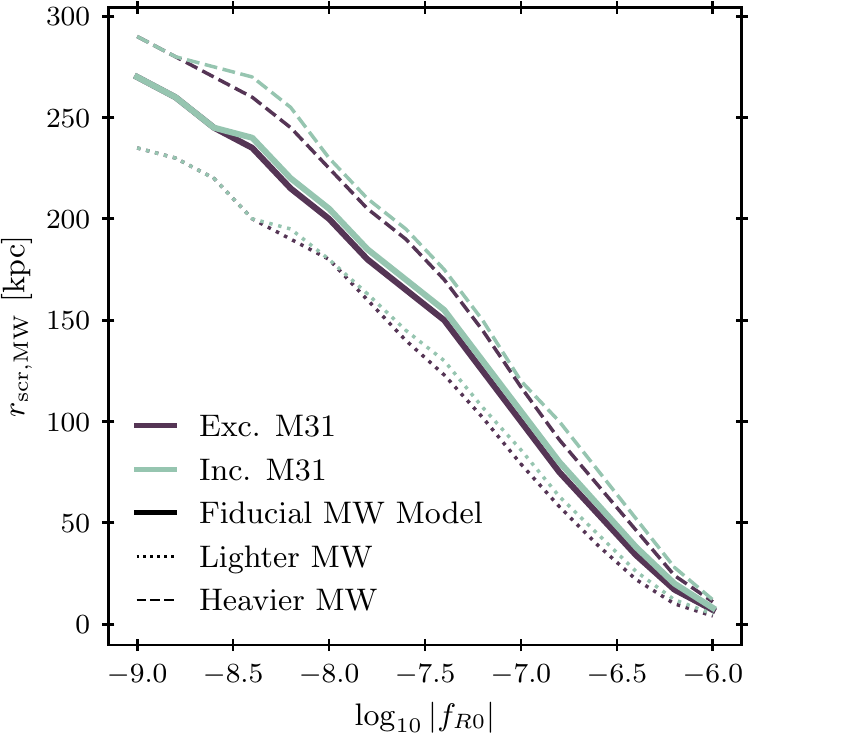}
    \caption{The disc-plane Milky Way screening radius as a function of $\log_{10}|f_{R0}|$. The solid lines show screening radii for the `fiducial' Milky Way model, i.e. the model described in Section \ref{S:Models:MW}. Meanwhile, the dotted and dashed lines represent Galaxy models in which the scale density $\rho_0$ of the dark matter halo has been rescaled by factors of 0.75 and 1.25 respectively. The colours of the lines indicate the environment around the Galaxy, where violet lines are for an isolated Milky Way model, while the green lines additionally incorporate the contribution of M31. In every case, the screening radius is calculated with \textsc{mg-gadget}.}
    \label{F:RscrfR0}
\end{figure}

\begin{figure*}[htp]
    \centering
    \includegraphics{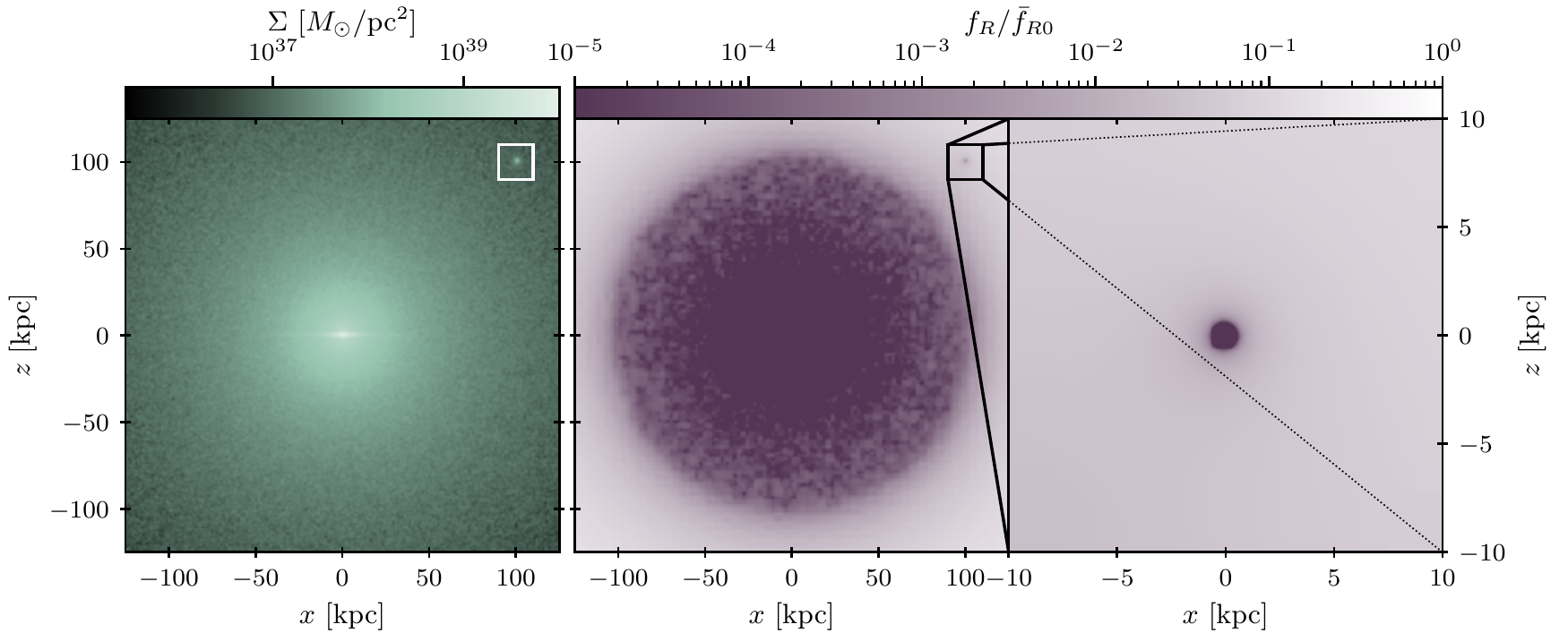}
    \caption{\textit{Left:} Edge-on particle density of Galaxy+satellite system fed to \textsc{mg-gadget} to calculate the scalar field profile. The location of the satellite is indicated by the inset box. \textit{Middle:} Scalar field profile for $f_{R0} = -10^{-7}$ across the same system. The Milky Way's screened region is clearly discernible, while the satellite also has a small central screened region, shown in the \textit{right-hand} panel, which shows a magnified image of the scalar field profile in a 20~kpc region centred around the satellite.}
    \label{F:MGGScreening}
\end{figure*}

\begin{figure*}[htp]
    \centering
    \includegraphics{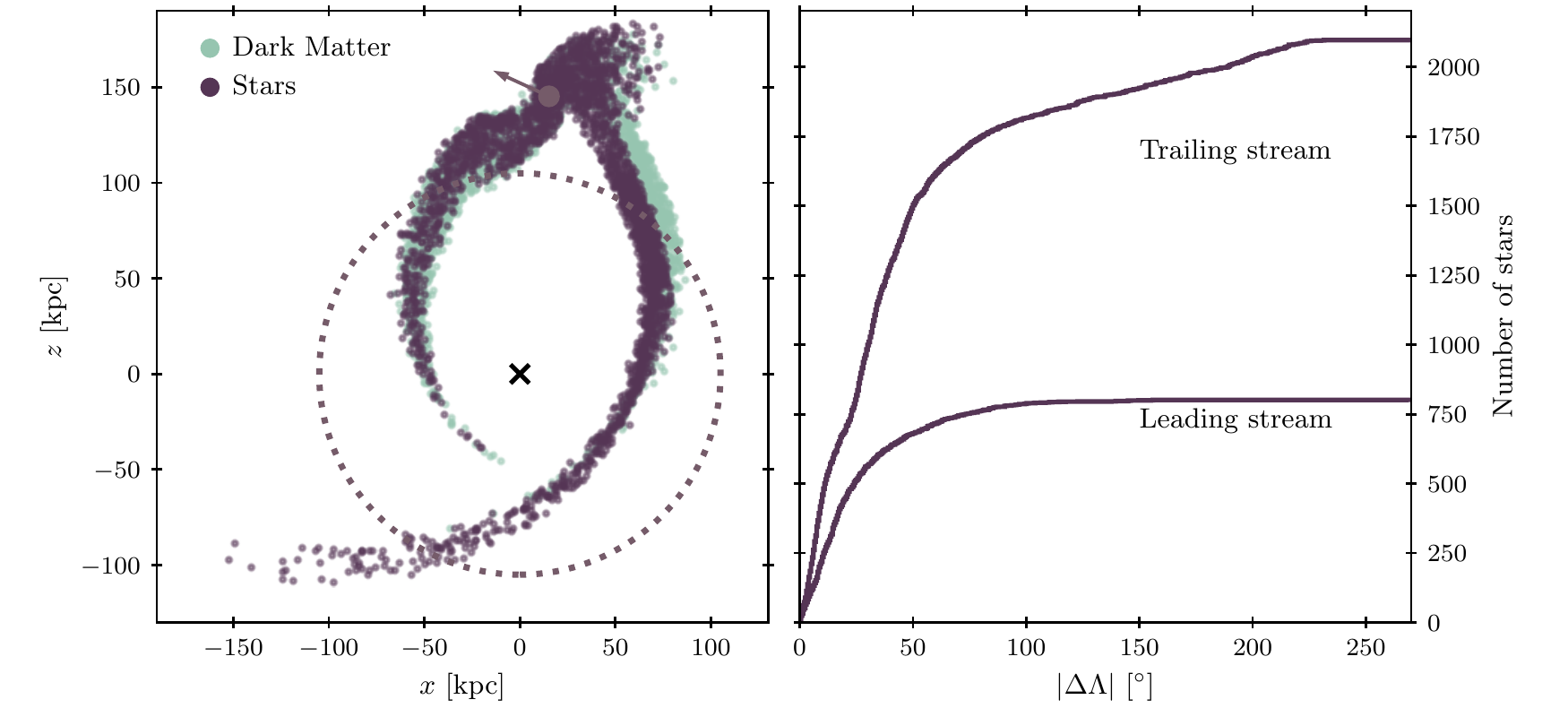}
    \caption{\textit{Left:} An image of a simulation of Satellite D, with $\rscrMW=105$ kpc, $\Qsat=0.8$, $\beta=0.4$. The dotted circle shows the location of the Milky Way screening radius, while the cross and filled circle show the locations of the Milky Way and satellite centres respectively. The arrow shows the current direction of motion of the satellite. \textit{Right:} Cumulative number of stars in either stream, as a function of longitude in the instantaneous orbital plane of the satellite. An Animation of this simulation is included in the Supplemental Material accompanying this article. This figure, taken together with Figure \ref{F:MGGScreening}, shows that $f(R)$ gravity with $f_{R0} \sim -10^{-7}$ should give a clear observational signature in stellar streams between 100 and 200 kpc.}
    \label{F:StreamCDF}
\end{figure*}

\section{Results}
\label{S:Results}

\subsection{Standard Gravity}

Figure \ref{F:StandardGravityImages} shows the images from the standard gravity simulations for all 4 satellites listed in Table \ref{T:StreamICs}. Each of the four quarters of the figure represents one of the satellites, as labelled in the top corner. The large subpanel in each quarter shows an image of the stream particles at the end of the simulation. As the stellar and dark matter particles are sampled from the same probability distribution initially (see assumption \ref{A:P1P2Profiles} in \S\ref{S:Methods}) and there is no EP-violation by a fifth force in these standard gravity simulations, the stars and dark matter particles are indistinguishable and are thus not plotted separately in this figure. The three smaller subpanels in each quarter show the average velocity along the stream, velocity dispersion along the stream, and velocity dispersion perpendicular to the stream, all as a function of stream longitude and all calculated in bins of particles along the stream. The bins are created adaptively, such that each bin contains 25 particles, including only the particles which have been stripped from the progenitor. Within each bin, the unit vector giving the direction \lq along the stream' is taken as the (normed) average velocity vector of all particles in the bin. This figure illustrates the diversity of our simulated streams, with a variety of morphologies and Galactocentric distances represented.

\subsection{Unscreened Fifth Force}

Turning to fifth forces, we first discuss results from an unscreened, EP-violating fifth force coupling only to dark matter ($\rscrsat=\rscrMW=0$). This is the case studied by \citet{Kesden2006a, Kesden2006b}. This case also applies in screened modified gravity with a (formally) universal coupling if stars self-screen, but screening is not triggered otherwise. In our work, the strength of the fifth force relative to gravity is given by $2\beta^2$, in keeping with the recent modified gravity literature, whereas Kesden and Kamionkowski used $\beta^2$. Thus, the simulation depicted in Figure \ref{F:TimeSeries} for example ($\beta=0.2,\ F_5/F_\mathrm{N}=0.08$), is most comparable to the `$\beta=0.3$' ($F_5/F_\mathrm{N}=0.09$) simulation in Refs.~ \cite{Kesden2006a,Kesden2006b}.

Figure \ref{F:OrbitalShapesBeta} shows the shape of Satellite B's orbit for a variety of values of $\beta$. In the absence of screening, the introduction of a fifth force as in Eq.~(\ref{E:FifthForce}) is tantamount to an overall linear rescaling of the Milky Way mass or gravitational constant by a factor of $1+2\beta^2$. As a consequence, the orbital period of the satellite is shorter and the apocentric distance smaller, as is apparent in Figure \ref{F:OrbitalShapesBeta}.

Figure \ref{F:TimeSeries} shows the positions of the dark matter and star particles in the simulation with $\rscrsat=\rscrMW=0$ and $\beta=0.2$ for Satellite C, at 11 equally spaced snapshots over time (recall that animations of selected simulations are available online).  The most striking feature is the asymmetry of the stellar stream. The preponderance of star particles populate the trailing stream, rather than the leading stream. The enhanced rotation speed of the satellite due to the fifth force means that the outward centrifugal acceleration of the stars outweighs the inward gravitational acceleration by the Milky Way. Consequently, stars are more likely to leave the satellite via the outer Lagrange point. Also, even some of the stars which are disrupted from the inner Lagrange point can eventually end up in the trailing stream, once sufficient time has passed for them to be overtaken by the satellite. Meanwhile, the dark matter particles experience the same fifth force as the satellite, and so there is (almost) no preferential disruption via either Lagrange point. The dark matter stream that forms, is consequently almost symmetric around the progenitor.

These effects are also apparent in Figure \ref{F:UnscreenedLongitudes}, which shows the longitude difference $\Delta\Lambda=\Lambda - \Lambda_\mathrm{sat}$ as a function of time for random subsamples of particles in the simulations without screening, with $\beta$ increasing in strength from 0.0 to 0.4 in steps of 0.1 for all 4 satellites. Here, $\Lambda$ is the longitude in the instantaneous orbital plane of the satellite and increases in the direction of the satellite's motion, so particles in the leading stream have positive $\Delta\Lambda$. The dark matter particles are stripped almost equally into the leading and trailing streams, leading to streams that are nearly symmetric about the progenitor for all values of $\beta$. For the star however, as $\beta$ increases, the particles are increasingly disrupted into negative longitudes, i.e. the trailing streams.

Sometimes, the satellite can be stripped completely of all of its stars. Then, the spatial separation between satellite and stream can be very large indeed, as no new stars become unbound from the satellite in order to bridge the gap. This occurs in Satellite A for both $\beta=0.3$ and $0.4$, as it loses all of its stars at its first pericentric passage. Satellite A, which is significantly less massive than our other satellites, does not have a sufficiently deep potential well for its stars to remain bound under the enhanced centrifugal force from the Milky Way. Some caution is needed because assumption \ref{A:ConstantSatPotential} for example (the assumption of an unchanging satellite mass and potential), may begin to break down when the disruption of the satellite due to the Milky Way is so severe. However, all of our satellites are, by assumption, dark matter dominated. Even in the simulations where the satellites lose all of their stars, they still retain a large fraction of their dark matter particles, and thus most of their assumed mass.

This result echoes a key finding of \citet{Keselman2009}, who argued that this prediction of stellar streams without associated progenitors could be related to the observed `orphan' streams of the Milky Way.

\subsection{Chameleon Screening}

We now show results from the chameleon simulations, i.e. the simulations with screening. Unlike the dark matter force investigated in the previous subsection, the fifth force here is universally coupled. However, as discussed in the Sections \ref{S:Introduction} and \ref{S:Theory}, an effective EP-violation arises because main sequence stars are self-screened against the fifth force in parameter regimes of interest.

Figure \ref{F:OrbitalShapesRscrMW} is the analogue of Figure \ref{F:OrbitalShapesBeta}, now showing the effect on the satellite's orbit of a varying Milky Way screening radius. In the case of the outermost screening radius of 45 kpc, nearly the entire orbit is situated within $\rscrMW$ and is therefore almost equivalent to the standard gravity case. Following along this orbit from plotted position of the progenitor, the other orbits peel away one by one, in order of increasing screening radius. In other words, once the orbit passes outside the screening radius, the fifth force becomes active and the orbit starts to diverge from the standard gravity case. Recalling from Eq.~(\ref{E:FifthForce}) that the fifth force is proportional to the mass between the test particle and the screening radius, the divergences do not become noticeable as soon as the orbit passes out of a given screening radius, but some time after, once this enclosed mass is large enough for an appreciable fifth force.

Looking instead at the impact of the Milky Way screening radius on stream asymmetries, one observable quantity is the ratio of the number of stars in the leading to the trailing stream,
\begin{equation}
\alpha =\frac{ N_\mathrm{lead}}{N_\mathrm{trail}}.
\end{equation}
Figure \ref{F:RscrMWComparison} shows this quantity as a function of Milky Way screening radius for all satellites, assuming $\Qsat=1$, i.e. fully unscreened satellites. To ensure a fair comparison between simulations, $\alpha$ is computed in each case at the moment of the satellite's third pericentric passage. As the MW screening radius increases, the asymmetry is progressively reduced. This appears to particularly be the case when $\rscrMW$ lies between the pericentre and apocentre of the orbit. This makes sense, as most tidal disruption occurs at and around pericentric passage. Therefore, screening the pericentre has the consequence of reducing the asymmetry of this disruption process. For all of our satellites, the streams are indistinguishable from those in the standard gravity case once $\rscrMW$ exceeds the apocentric distance.

We have observed in our simulations interesting signatures of chameleon gravity other than the stellar asymmetry. Examples of these are depicted in Figure \ref{F:InterestingSignatures}. First, in the extreme (high $\beta$) fifth force regime, the orbital paths of released stars around the Milky Way differ appreciably from their progenitor. However, because stars are released from the progenitor at different times, this also means that the liberated stars can be on different Milky Way orbits from each other. If most releases occur at pericentric passages, this can lead to a `striping' effect, with neighbouring undulations of stars on the sky, corresponding to streams of stars released at successive pericentric passages. This effect is visible in the upper panel of Figure \ref{F:InterestingSignatures}.

Secondly, if the satellite itself is fully screened or almost so (i.e. low $\Qsat$), then it orbits the Milky Way more slowly than the dark matter that has been released and inhabits unscreened space. Then, we observe the opposite asymmetry to that of the stars: the dark matter is preferentially disrupted into the leading stream rather than the trailing stream. This effect is shown in the lower panel of Figure \ref{F:InterestingSignatures}. While interesting, this effect is of course not readily accessible to observations.

\subsection{Future Constraints}
\label{S:Results:Future}

The later Gaia data releases will likely enable the discovery of stellar streams at large distances from the Galactic centre. As shown in Figure \ref{F:RscrMWComparison}, such streams are able to probe larger Milky Way screening radii, and therefore `weaker', or more screened, regions of modified gravity parameter space.

Figure \ref{F:DAsymmetry} shows $\alpha$ evaluated for all of our simulations of satellite D, as a function of $\rscrMW$, $\Qsat$, and $\beta$. As with Figure \ref{F:RscrMWComparison}, $\alpha$ is computed in each simulation at the moment of the satellite's third pericentric passage. This figure illustrates many of our earlier points; increasing $\beta$ increases the magnitude of the asymmetry, but the asymmetry is reduced by increasing $\rscrsat$ (reducing $\Qsat$) or $\rscrMW$. In the $\beta=0.4$ case, approximately comparable to $f(R)$ gravity, the asymmetries grow large when $\rscrMW \lesssim 100$ kpc, assuming the satellite is fully unscreened ($\Qsat=1$). Notably, this lies between the apocentre and pericentre of the satellite's orbit. Most tidal disruption occurs at pericentric passage, but here there is still enough disruption outside the screening radius, and sufficient numbers of leading stars lagging behind the satellite, that a large asymmetry develops anyway.

We can again use Hu-Sawicki $f(R)$ gravity to give an indication of the kinds of constraints attainable here. Figure \ref{F:RscrfR0} shows how the Milky Way screening radius depends on the parameter $f_{R0}$. These calculations were performed using the scalar field solver within the $f(R)$ N-body code \textsc{mg-gadget} \citep{Puchwein2013}. \textsc{mg-gadget} uses a Newton-Gauss-Seidel relaxation method to solve the $f(R)$ equations of motion, calculating the scalar fields and fifth forces everywhere across a given mass distribution or within a given simulation volume. Such methods were first explored in the work of \citet{Oyaizu2008}, and the subsequent years have seen a proliferation of codes simulating a myriad of modified gravity cosmologies \citep{Li2009, Schmidt2009, Zhao2010, Brax2011, Brax2012, Li2012, Li2013, Llinares2013}.

Given a mass model for the Galaxy, this can then be sampled with a large number of particles, which are in turn fed to \textsc{mg-gadget} to calculate the scalar field profile---and thus the Galaxy screening radius---for a given value of $f_{R0}$. In each case, we truncate the mass model at $r_{100}$, the radius encompassing a region with density 100 times the cosmic critical density. In the case of the Milky Way, this radius has been shown to lie close to the `splashback radius', a reasonable definition for the edge of the Galaxy's halo \citep{Deason2020}.

The solid purple line in Figure \ref{F:RscrfR0} shows the screening radii for the `fiducial' Milky Way model described in \S\,\ref{S:Models:MW}, i.e. the model used throughout our simulations. This illustrates the sensitivity of stream asymmetries as a probe of chameleon gravity.

However, the overall mass of the Milky Way is highly uncertain, and the primary source of uncertainty is the dark matter halo, particularly in its outer regions \citep{Bland2016}. Thus, it is interesting to explore how this predicted sensitivity depends on the overall mass of the Galaxy. The radius $r_{100}$ for our fiducial model is $\sim$300 kpc, and the total mass $M_{100}$ (including baryons) enclosed within this radius is approximately $1.5 \times 10^{12} M_\odot$. Figure \ref{F:RscrfR0} additionally shows screening radii calculated for less massive (dotted) and more massive (dashed) Milky Way models. In these models the scale density $\rho_0$ of the dark matter halo has been rescaled by factors of 0.75 and 1.25 respectively. These rescalings still lead to sensible values for the halo concentration, and correspond to masses of $M_{100} = 1.1$ and $2.0 \times 10^{12} M_\odot$ (note the overall masses are not rescaled by exactly 0.75 and 1.25, because $r_{100}$ changes along with the density normalisation). As expected, at fixed $f_{R0}$, increasing (decreasing) the mass leads to an expansion (reduction) of the screening radius. These masses bracket a large range of reasonable estimates for the Milky Way mass, and so the region of the figure enclosed by these lines should in principle include the `true' screening radius of the Milky Way in an $f(R)$ Universe.

One additional caveat, however, is that this treatment has ignored the environmental contributions to the scalar field by the Local Group. As a first approximation of this effect, the green lines in Figure \ref{F:RscrfR0} show the Milky Way screening radii when a mass distribution for M31 is added to the \textsc{mg-gadget} input. The model used for M31 is identical to our fiducial Milky Way model, centred at $(-380,620,-280)$~kpc in Galactocentric coordinates \citep{Marel2012}. There is a systematic upward shift in the Milky Way screening radii at all $f_{R0}$ values except at $\log_{10}|f_{R0}| \lesssim -8.5$, where $r_{MW-M31} > \lambda_C$. However, the magnitude of this shift is typically rather small, on the order of a few kpc. This may increase with a more sophisticated model of the Local Group incorporating M33 and various other smaller galaxies, as well as the smooth intervening mass distribution. However, as the Milky Way and M31 are by far the most massive members of the Local Group, it seems likely that environmental screening will remain a subdominant effect.

We see from Figure \ref{F:RscrfR0} that Satellite D is able to probe the region $\log_{10}|f_{R0}| \gtrsim -7.2$. However, if the satellite itself is partially screened, the sensitivity is greatly reduced. It is natural therefore to wonder about the degree to which a satellite would be screened at these values of $f_{R0}$ and this region of the Milky Way's halo.

Figure \ref{F:MGGScreening} shows the scalar field profile around the Milky Way for $f_{R0} = -10^{-7}$, again inferred using \textsc{mg-gadget}. A Hernquist sphere identical to Satellite D has been inserted at Galactocentric $(X=100, Y=0, Z=100)$ kpc. There is a clear screened region in the centre of the Milky Way halo, with $\rscrMW\approx100$ kpc. The satellite is also partially screened, with a screened region of $\rscrsat\approx0.6$ kpc at its centre. This corresponds to $\Qsat=0.8$. Comparing to Figure \ref{F:DAsymmetry}, the suggestion is that in an $f(R)$ Universe, this satellite would provide very asymmetric streams. This is demonstrated in Figure \ref{F:StreamCDF}, which shows a simulation with a similar setup: Satellite D with $\beta=0.4$, $\rscrMW=105$ kpc, and $\Qsat=0.8$. The left-hand panel shows the stream, while the right-panel shows a more sophisticated observable signature than the asymmetry parameter: the cumulative number function of stars in each stream as a function of longitude in the orbital plane of the satellite. The difference in the two curves is rather striking, and should be clearly discernible in the data.

The examples shown in Figures \ref{F:MGGScreening} and \ref{F:StreamCDF} serve as proof of concept, demonstrating that stellar streams in the outer reaches of our Galaxy's halo are a sensitive probe of modified gravity. The observation of highly symmetric streams at large Galactocentric distances would rule out sizeable fifth forces that couple differently to dark matter and stars in the outskirts of the Milky Way. This in turn would provide sensitive constraints on screened modified gravity theories. For instance, looking at Figure \ref{F:RscrfR0}, symmetric streams at distances of $\sim 150-300$ kpc would require $|f_{R0}| \sim 10^{-7.5}$ or even $10^{-8}$ to avoid sizeable fifth forces in that radial range. This would be among the tightest constraints on $f(R)$ gravity achievable to date. However, we caution that environmental screening of the satellite may play a more significant role at these levels, but Figure \ref{F:DAsymmetry} suggests that only if the satellite is fully screened does the signal disappear entirely. Even when $\Qsat=0.1$, i.e. 90\% of the mass is screened, there is still an appreciable signal. So, given the observation of a large number of symmetric streams, and if there is little environmental screening by the Local Group, then constraints down to these levels are feasible.

On the other hand, observations of highly asymmetric streams would strengthen the case for screened modified gravity theories. It should be noted, however, that mild asymmetries can arise due to dynamical effects. Indeed, an asymmetry between the leading and trailing streams is expected from Eq.~(\ref{E:natasym}). This may be compounded by dynamical interactions with dark subhaloes or other satellites~\cite{Erkal2015}, asymmetries in the stellar populations in the progenitor satellite~\cite{Penarrubia2010,Bonaca2019}, effects of the Galactic bar~\cite{Hattori2016, Erkal2017, Pearson2017} and regions of chaos in the Galactic potential~\cite{Price2016}. Such effects would have to be carefully weighed before a modified gravity interpretation could be seriously considered for such observations.

\section{Conclusions}
\label{S:Conclusions}

We have investigated possible imprints of chameleon gravity on stellar streams from dwarf galaxies around the Milky Way. While canonical chameleon theories are universally coupled, an effective violation of the equivalence principle (EP) arises because of the self-screening of main sequence stars, as noted by \citet{Hui2009}. Consequently, stars are preferentially stripped from the progenitor into the trailing stream rather than the leading stream.

We have created a restricted N-body code (\texttt{smoggy}; made publicly available \footnote{\url{https://github.com/aneeshnaik/smoggy}}), and used it to simulate the formation of tidal streams from progenitors with a variety of masses and Galactocentric distances. We considered a range of modified gravity scenarios (coupling strength, Milky Way screening level, satellite screening level) in each case.

As found by \citet{Kesden2006a,Kesden2006b}, an EP-violating fifth force that couples to dark matter but not baryons causes asymmetries to develop in stellar streams with dark matter-dominated progenitors. The stars are preferentially disrupted via the outer Lagrange points into the trailing streams. We have corrected and augmented the analytic calculations of Ref.~\citep{Kesden2006a} for point masses so that they are also applicable to extended Galactic mass distributions like isothermal spheres. The effect of these changes is to make the test more sensitive to EP-violating fifth forces. For the most massive dwarf spheroidals, like the Sagittarius or Fornax, the criterion given in Eq.~(\ref{E:gencaseasym}) suggests values of $\beta^2 \gtrsim 10^{-3}$ can be probed. For the smallest dwarf spheroidals such as Segue 1 with a mass of  $6 \times 10^5 \Msun$, then values of $\beta^2 \gtrsim 10^{-4}$ are in principle accessible. As a rule of thumb for a satellite with mass $\Msat$ at a location enclosing a Milky Way mass $\MMW$, the form of the criterion suitable for a flat rotation curve galaxy is
    \begin{equation}
\label{E:threebodycritconcl}
    \beta^2 \gtrsim 2^{-5/3} \left( \frac{\Msat}{\MMW} \right)^{2/3}.
\end{equation}
This asymmetry also occurs in the chameleon context, when screening radii are introduced to the Milky Way and satellite, and with stars self-screening. The magnitude of the asymmetry depends on the coupling strength $\beta$, the Milky Way screening radius, as well as the degree of screening of the stream progenitor; large values of $\beta$ give large asymmetries, but these are reduced with increasing $\rscrMW$ and $\rscrsat$.

Our simulations -- the most comprehensive to date for the formation of tidal streams under chameleon gravity -- have revealed further interesting effects. First, the trailing stellar stream may become detached from the dark matter progenitor if all the stars are exhausted by earlier pericentric stripping (cf. \cite{Keselman2009}). As an example, this effect is visible in Figure~\ref{F:UnscreenedLongitudes} and occurs for low mass satellites in the extreme fifth force regime. Second, prominent striations in the stellar trailing tail may exist if stars are stripped at repeated pericentric passages by a strong fifth force. Thirdly, if the satellite is fully screened, its orbital frequency is lower than that of its associated dark matter. This leads to strong asymmetries in the dark matter distribution, which is preferentially liberated into the leading tidal tail.

Taking Hu-Sawicki $f(R)$ gravity with $f_{R0} = -10^{-7}$ as an example, we derive a Milky Way screening radius of around 100 kpc. A massive dwarf spheroidal galaxy at a distance of $\approx 150$ kpc -- such as Fornax -- would be partially screened but would nonetheless produce highly asymmetric streams under tidal disruption.

The ratio of the cumulative number function of stars in the leading and trailing stream as a function of longitude from the satellite is computable from simulations, measurable from the observational data and can provide a direct test of theories with screening mechanisms like chameleon gravity. The later Gaia data releases may lead to discoveries of stellar streams at distances $\gtrsim 100$ kpc from the Galactic centre. These streams will be a sensitive probe of modified gravity; such highly asymmetric streams at these distances would be tell-tale signatures of modified gravity.

On the other hand, if the data uncover a number of very symmetric streams, then constraints down to the level of $f_{R0} \sim -10^{-8}$---the tightest constraints to date---could be attainable if the screening of the satellite and other nuisance parameters are carefully taken into account. Also, our assumption that the Compton wavelength is much larger than relevant length scales begins to break down at such values of $f_{R0}$, and Yukawa suppression will become appreciable below $f_{R0} \sim -10^{-8}$. Of course, the investigation need not be limited to Hu-Sawicki $f(R)$ gravity. Sensitive constraints will also be attainable in the general chameleon parameter space, and we merely use $f(R)$ gravity as a fiducial theory.

This desirability of streams at large distances suggests another interesting avenue for exploration: stellar streams around other galaxies. Streams have already been observed around other galaxies (e.g. \citep{Martinez2008, Martinez2010}), and it seems likely that future wide-field surveys such as LSST \citep{Ivezic2019} will uncover more streams at large distances from their host galaxies. This, combined with a calculation of the host galaxy screening properties (e.g. via the screening maps of Ref.~\citep{Desmond2018c}) could also be a sensitive probe of screened modified gravity.

Finally, we note that other screened modified gravity theories can also be probed with stellar streams. For instance, the symmetron screening mechanism \citep{Hinterbichler2010,Hinterbichler2011} has a simple density threshold as a screening criterion. Consequently, there will necessarily be a region of parameter space in which the stars are screened, but the surrounding diffuse dark matter component is not. In this regime, stream asymmetries will also be present and are worthy of future investigation.

\begin{acknowledgments}
The authors would like to acknowledge Matt Auger, Vasily Belokurov, Sergey Koposov, Jason Sanders, and Denis Erkal for helpful discussions, and the anonymous referee for reviewing and suggesting various improvements to the manuscript.

APN  thanks  the  Science  and  Technology  Facilities  Council (STFC) for their PhD studentship. HZ acknowledges support by the Kavli Foundation. ACD acknowledges partial support from STFC under grants ST/L000385 and ST/L000636. This work used the DiRAC Data Analytic system at the University of Cambridge, operated by the University of Cambridge High Performance Computing Service on behalf of the STFC DiRAC HPC Facility (www.dirac.ac.uk). This equipment was funded by BIS National E-infrastructure capital grant (ST/K001590/1), STFC capital grants ST/H008861/1 and ST/H00887X/1, and STFC DiRAC Operations grant ST/K00333X/1. DiRAC is part of the National E-Infrastructure.

\end{acknowledgments}

\bibliography{library}

\end{document}